\documentclass[sigplan,screen]{acmart}
\renewcommand\footnotetextcopyrightpermission[1]{} 

\usepackage{algorithm}
\usepackage[noend]{algpseudocode}
\usepackage{amsmath}
\usepackage{listings}
\usepackage{xcolor}
\usepackage[normalem]{ulem}
\usepackage{multirow}
\usepackage{enumitem}

\mathchardef\mhyphen="2D

\definecolor{codegreen}{rgb}{0,0.6,0}
\definecolor{codegray}{rgb}{0.5,0.5,0.5}
\definecolor{codepurple}{rgb}{0.58,0,0.82}

\lstdefinestyle{mystyle}{
    frame=tb,
    commentstyle=\color{codegreen},
    keywordstyle=\color{magenta},
    numberstyle=\scriptsize,
    stringstyle=\color{codepurple},
    basicstyle=\ttfamily\small,
    breakatwhitespace=false,
    breaklines=true,
    captionpos=b,
    keepspaces=true,
    numbers=left,
    numbersep=5pt,
    showspaces=false,
    showstringspaces=false,
    showtabs=false,
    tabsize=2
}

\newcommand{\tabincell}[2]{\begin{tabular}{@{}#1@{}}#2\end{tabular}}

\AtBeginDocument{%
  }

\begin{document}
\pagestyle{plain}

\title{HIDA: A Hierarchical Dataflow Compiler for High-Level Synthesis}

\author{Hanchen Ye}
\orcid{0000-0002-6646-8146}
\affiliation{%
    \institution{University of Illinois Urbana-Champaign}
    \country{United States}
}
\email{hanchen8@illinois.edu}

\author{Hyegang Jun}
\orcid{0000-0002-7879-6884}
\affiliation{%
    \institution{University of Illinois Urbana-Champaign}
    \country{United States}
}
\email{hgjun2@illinois.edu}

\author{Deming Chen}
\orcid{0000-0002-3016-0270}
\affiliation{%
    \institution{University of Illinois Urbana-Champaign}
    \country{United States}
}
\email{dchen@illinois.edu}

\renewcommand{\shortauthors}{Hanchen Ye, Hyegang Jun, and Deming Chen}

\begin{abstract}
Dataflow architectures are growing in popularity due to their potential to mitigate the challenges posed by the memory wall inherent to the Von Neumann architecture. At the same time, high-level synthesis~(HLS) has demonstrated its efficacy as a design methodology for generating efficient dataflow architectures within a short development cycle. However, existing HLS tools rely on developers to explore the vast dataflow design space, ultimately leading to suboptimal designs. This phenomenon is especially concerning as the size of the HLS design grows. To tackle these challenges, we introduce \emph{HIDA}\footnote{\url{https://github.com/UIUC-ChenLab/ScaleHLS-HIDA}}, a new scalable and hierarchical HLS framework that can systematically convert an algorithmic description into a dataflow implementation on hardware. We first propose a collection of efficient and versatile dataflow representations for modeling the hierarchical dataflow structure. Capitalizing on these representations, we develop an automated optimizer that decomposes the dataflow optimization problem into multiple levels based on the inherent dataflow hierarchy. Using FPGAs as an evaluation platform, working with a set of neural networks modeled in PyTorch, HIDA achieves up to 8.54$\times$ higher throughput compared to the state-of-the-art~(SOTA) HLS optimization tool. Furthermore, despite being fully automated and able to handle various applications, HIDA achieves 1.29$\times$ higher throughput over the SOTA RTL-based neural network accelerators on an FPGA.
\end{abstract}




\maketitle

\thispagestyle{empty}

\section{Introduction}
\label{sec:introduction}

With the decline of Moore’s law, it is no longer possible to expect the price of computation to decrease year over year.
As a result, \emph{customized} and \emph{domain-specific} accelerators are becoming well accepted in combating the physical limitations of silicon, including those implemented on ASICs~\cite{du2015shidiannao, jouppi2017datacenter, genc2021gemmini} and reconfigurable platforms, such as FPGAs~\cite{zhang2021boostgcn, zhou2021mocha, ye2020hybriddnn}. Historically, the cost of developing hardware accelerators has always remained astronomically high. In this context, high-level synthesis~(HLS) is a promising solution that can \textit{synthesize} high-level algorithmic description to a hardware description language~(HDL) implementation~\cite{cong2011high}.

\textbf{Dataflow Architecture.}
An important computation architecture for customized hardware accelerators is \emph{dataflow}, which enables the parallel temporal execution of multiple coarse-grained tasks~\cite{moreau2018vta, prabhakar2017plasticine, zhang2020dnnexplorer}. Unlike the Von Neumann architecture that constantly grapples with the memory wall, dataflow architecture can exploit the on-chip communication between tasks to avoid frequent external memory access. As long as an application is dataflow feasible, a well-designed dataflow architecture can efficiently execute the application with reduced power and bandwidth utilization~\cite{umuroglu2017finn, wei2018tgpa, fahim2021hls4ml}.

\textbf{Existing Dataflow Approaches.}
Commercial HLS tools typically provide programming interfaces for users to implement dataflow structures, such as the AMD Vitis HLS \texttt{dataflow} directive~\cite{vitishls2022userguide}, Intel HLS \emph{system of tasks}~\cite{intelhls2022userguide}, and LegUp \texttt{thread} APIs~\cite{legup2021document}. However, it is still difficult to implement a dataflow-oriented HLS design with sequential languages, such as C/C++. Therefore, academic HLS tools have pushed for approaches that decouple algorithm specification from hardware customizations in compute, data types, and memory~\cite{ben2019stateful, lai2019heterocl, xiang2022heteroflow, ikarashi2022exocompilation}, or introduce specialized HLS primitives~\cite{margerm2018tapas, koeplinger2018spatial, chi2021extending, guo2022tapa}. These approaches have effectively improved productivity and quality compared to industrial HLS tools. Note that there also exist recent frameworks~\cite{hpca2022scalehls, ejjeh2022hpvm2fpga} that can automatically generate dataflow design without manual code rewriting. However, these automated tools cannot systematically model dataflow architectures, limiting them to the generation of suboptimal simple designs.

\textbf{Unexplored Opportunities.}
Although existing HLS tools can enable dataflow designs, they still heavily rely on the user to make the hard design decisions, including but not limited to parallelization strategy, tiling strategy, memory hierarchy, data layout, etc. More importantly, the design spaces of different tasks in the dataflow are tightly coupled with each other due to two reasons: (1)~an efficient dataflow architecture demands the latency to be balanced  across different tasks as the critical task determines the overall achievable performance; (2)~the inter-task communications are often established through streaming channels or on-chip buffers instead of hierarchical shared memory. Meanwhile, large-scale dataflow often gravitates towards a \emph{hierarchical} structure, as dataflow tasks are naturally represented by nested graphs, further complicating the design space.

As a result, the vast design space can prohibit programmers from reasoning about various design choices and finding the optimized design point. This can eventually lead to non-ideal performance and efficiency, thereby thwarting the promise of existing dataflow approaches. We observed that many HLS-augmentation tools have proposed DSE engines using different algorithms, including polyhedral techniques~\cite{zuo2015polyhedral, zuo2017accurate, agostini2022mlir, zhao2022polsca}, graph analysis~\cite{zhong2016lin, zhao2017comba, huang2021pylog, dac2022scalehls}, and machine learning~\cite{sohrabizadeh2022autodse, yu2021chimera, ejjeh2022hpvm2fpga, jun2023autoscaledse}. These tools can effectively explore the local design space of a single task or kernel. However, they cannot handle the dataflow-oriented exploration of multiple tasks due to the inter-task coupling and the complicated dataflow hierarchy.

\textbf{HIDA Approach.}
With the discussion above, we concluded that the challenges presented in the design and optimization of dataflow architecture \textit{cannot} be fully addressed by existing HLS approaches, which rely on programmers to explore the vast design space manually. We argue that compilers will and should play an important role in the design process - the hierarchical characteristics of dataflow architecture should be systematically represented and modeled, on which an optimization pipeline should be built to handle the inter- and intra-task optimizations comprehensively.

Under this mantra, we propose \emph{HIDA}, an HLS framework with \underline{hi}erarchical \underline{da}taflow intermediate representations~(IR) and optimizations, enabling the automated transformation of algorithmic hardware descriptions to efficient dataflow architectures. The main contributions of HIDA are as follows:
\begin{itemize}[leftmargin=*]
    \item We propose a new dataflow IR called HIDA-IR that models dataflow at two different levels of abstraction, \textit{Functional} and \textit{Structural}, to capture the dataflow characteristics and multi-level hierarchy, enabling effective optimizations.
    \item We propose a new dataflow optimizer called HIDA-OPT, featuring a pattern-driven task fusion algorithm and an intensity- and connection-aware dataflow parallelization algorithm geared toward maximum efficiency.
    \item We enable an end-to-end and extensible compilation stack supporting PyTorch and C++ inputs, empowering the user to rapidly experiment with various design parameters and prototype new dataflow architectures.
    \item We perform comprehensive FPGA evaluations of HIDA. On a set of neural networks, HIDA achieves 8.54$\times$ and 1.29$\times$ higher throughputs over the SOTA HLS optimization framework and RTL-based neural network accelerator.
\end{itemize}


\begin{figure*}[t]
\centering
\begin{minipage}{0.39\textwidth}
    \centering
    \captionof{table}{LeNet accelerator design. \emph{CPF} and \emph{KPF} denote the channel and kernel parallel factor.}
    \label{tab:lenet_model}
    \vspace{-4pt}
    \small
    \setlength\tabcolsep{4pt}
    \begin{tabular}{cccc}
        \toprule
        \textbf{Layer} & \textbf{Task} & \textbf{Factor} & \textbf{Range} \\
        \midrule
        (All Layers) & - & $BATCH$ & \{1, 5, 10, 15, 20\} \\
        \midrule
        Conv+ReLU & \multirow{2}{*}[-2pt]{Task1} & \multirow{2}{*}[-2pt]{$KPF_{task1}$} & \multirow{2}{*}[-2pt]{\{1, 2, 3, 6\}} \\
        \cmidrule(lr){1-1}
        Pool & \\
        \midrule
        Conv+ReLU & \multirow{2}{*}[-3pt]{Task2} & \multirow{2}{*}[-3pt]{\tabincell{c}{$KPF_{task2}$ \\ $CPF_{task2}$}} & \multirow{2}{*}[-3pt]{\tabincell{c}{\{1, 2, 4, 8, 16\} \\ \{1, 2, 3, 6\}}} \\
        \cmidrule(lr){1-1}
        Pool & \\
        \midrule
        Conv+ReLU & Task3 & \tabincell{c}{$KPF_{task3}$ \\ $CPF_{task3}$} & \tabincell{c}{\{1, 2, 3, 4, 6, 8\} \\ \{1, 2, 4, 8, 16\}} \\
        \midrule
        Linear & Task4 & - & - \\
        \bottomrule
    \end{tabular}
    \vspace{10pt}
    \captionof{table}{Evaluation results of LeNet.}
    \label{tab:lenet_result}
    \vspace{-4pt}
    \small
    \setlength\tabcolsep{3pt}
    \begin{tabular}{cccc}
        \toprule
        & \textbf{Expert} & \textbf{Exhaustive} & \textbf{HIDA} \\
        \midrule
        \textbf{Resource Util.} & 95.5\% & 99.2\% & 95.0\% \\
        \textbf{Throu. (Imgs/s)} & 41.6k & 49.9k & 53.2k \\
        \textbf{Develop Cycle} & 40 hours & 210 hours & 9.9 mins \\
        \bottomrule
    \end{tabular}
\end{minipage}
\hfill
\begin{minipage}{0.59\textwidth}
    \centering
    \includegraphics[width=\textwidth]{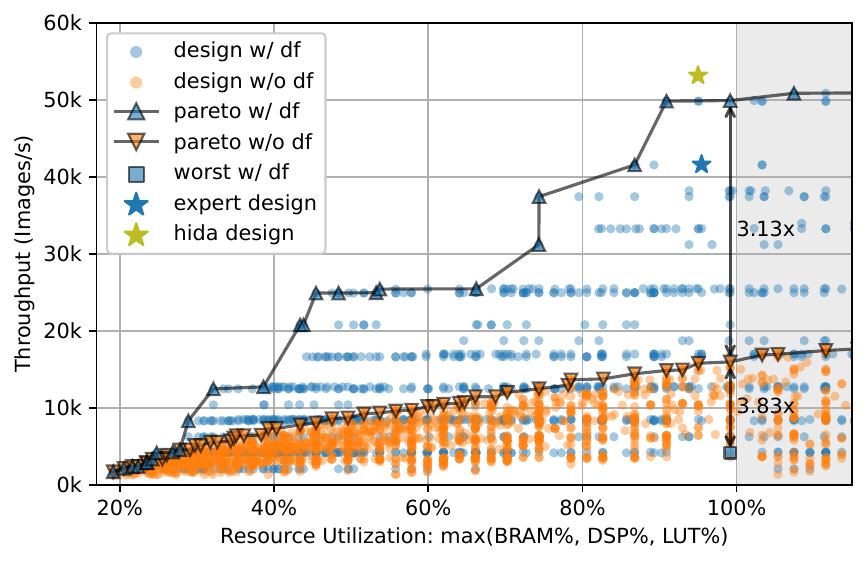}
    \vspace{-17pt}
    \captionof{figure}{Exhaustive design space search of the LeNet accelerator. \emph{w/ df} and \emph{w/o df} indicate whether dataflow is enabled. \emph{Worst w/ df} is the dataflow design with the worst quality. \emph{Expert design} is hand-tuned by HLS experts. \emph{HIDA design} is automatically generated from HIDA.}
    \label{fig:lenet_search}
\end{minipage}
\end{figure*}

\section{Motivation}
\label{sec:motivation}

Due to the inherent disjoint between the Von Neumann-centric programming model and the dataflow programming model, it is easy for HLS designs to leave a large portion of the achievable performance on the table. To better understand the challenges presented in existing HLS tools, we implemented an HLS-based LeNet~\cite{lecun1998gradient} accelerator on an AMD PYNQ-Z2 FPGA as a case study.

\textbf{Design Process.}
Table~\ref{tab:lenet_model} shows the structure of the LeNet model, which consists of 6 layers in total. We followed the steps below to design the HLS-based accelerator:
\begin{enumerate}[leftmargin=*]
    \item We rewrote the LeNet model in C++ as the baseline design, which is also used for the testbench and simulation in all subsequent steps. (2 hours)
    \item We applied \emph{layer fusion} and \emph{parallelization} to the baseline design following the strategy summarized in Table~\ref{tab:lenet_model}. The listed parallel factors are selected with heuristics~\cite{zhang2015optimizing} and implemented with manual loop tiling and loop \texttt{unroll} directive insertion. (10 hours)
    \item We rewrote the design to enable coarse-grained \emph{dataflow} and loop \emph{pipeline}. Specifically, we outlined all tasks, implemented off-chip memory interfaces, and partitioned inter-task on-chip buffers with heuristics~\cite{wang2013memory}. (8 hours)
    \item We iterated on different settings of parallel factors and directive configurations by rewriting and evaluating the design in AMD Vitis HLS until we were satisfied with the quality of the results. (20 hours)
\end{enumerate}
Overall, we spent around 40 hours designing and fine-tuning the HLS-based LeNet accelerator. Then, we parameterized all six parallel factors listed in Table~\ref{tab:lenet_model} and developed a TCL script exhaustively traversing each configuration under both dataflow and non-dataflow settings. This took another 170 hours. Finally, as a comparison, we automatically generated an HIDA-based design, which took 0.4 minutes to compile and 9.5 minutes for AMD Vitis HLS to generate RTL.

\textbf{Results and Analysis.}
Figure~\ref{fig:lenet_search} shows the exhaustive search results of the LeNet accelerator in the throughput-resource space. Table~\ref{tab:lenet_result} summarizes the evaluation results and development cycles of the expert design, the best design from the exhaustive search, and the HIDA design. The key observations are summarized as follows:
\begin{itemize}[leftmargin=*]
    \item \emph{Dataflow designs are Pareto-dominating.} We can clearly observe a large throughput/resource gap between the Pareto frontiers with and without dataflow. The best dataflow design achieves 3.13$\times$ higher throughput than the non-dataflow counterpart under the same resource constraints.
    \item \emph{Dataflow cannot guarantee a good trade-off.} We can observe tons of dataflow designs dominated by non-dataflow designs. Under the same resource constraints, the best non-dataflow design achieves 3.83$\times$ higher throughput than the dataflow design with the worst quality.
    \item \emph{Dataflow design space is vast.} In the layer fusion, spatial parallelization, and array partition steps, we have pruned a large amount of design points based on heuristics. However, the resulting design space still contains more than $2.4\times10^4$ points and costs hundreds of CPU hours to search exhaustively, owing to the fact that each design point takes 2-10 minutes for Vitis HLS to evaluate.
    \item \emph{Dataflow design space is difficult to comprehend.} In Table~\ref{tab:lenet_result}, we can observe that the \emph{exhaustive} design achieves 1.20$\times$ higher throughput than the hand-tuned \emph{expert} design. We attribute this to the inter-task design space coupling. The complicated dataflow design space makes it substantially difficult to find the optimized design point by reasoning about the trade-off empirically.
    \item \emph{Automated tool outperforms exhaustive search.} HIDA further improves the throughput of the \emph{exhaustive} design by 1.06$\times$. After comparing the two designs, we found HIDA to have automatically explored additional parallelizable dimensions apart from the six in the \emph{Factor} column of Table~\ref{tab:lenet_model}, such as the feature map width and height dimensions, presenting a path that can increase the design efficiency further. This indicates that using heuristics suitable for single-task or non-dataflow designs may accidentally prune away valuable design points for dataflow designs.
\end{itemize}

\textbf{Need for Scalable Automation Tools.}
For the LeNet case study, while the expert design took tens of hours to develop and ended up with a sub-optimal design, HIDA only took minutes to generate the design and achieved the best quality of results. One should expect that as the complexity and size of the target design increase, the development time will grow dramatically, while the manual design quality will decrease due to the vast and complicated dataflow design space.
In summary, \emph{productivity}, \emph{performance}, and \emph{scalability} problems of dataflow architecture are the three strong motivators for a scalable HLS optimization tool.

\begin{figure}
    \centering
    \includegraphics[width=\linewidth]{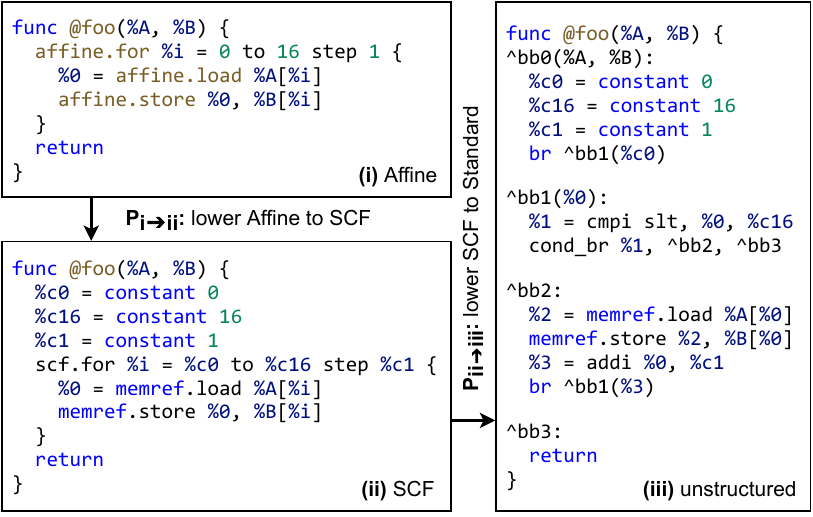}
    \vspace{-17pt}
    \caption{An IR example. \texttt{affine} and \texttt{scf} dialect are structured control flow IRs that can be lowered to unstructured IR. All types are omitted for simplicity.}
    \label{fig:mlir}
\end{figure}

\begin{figure*}[t]
\begin{minipage}{0.55\textwidth}
    \includegraphics[width=\textwidth]{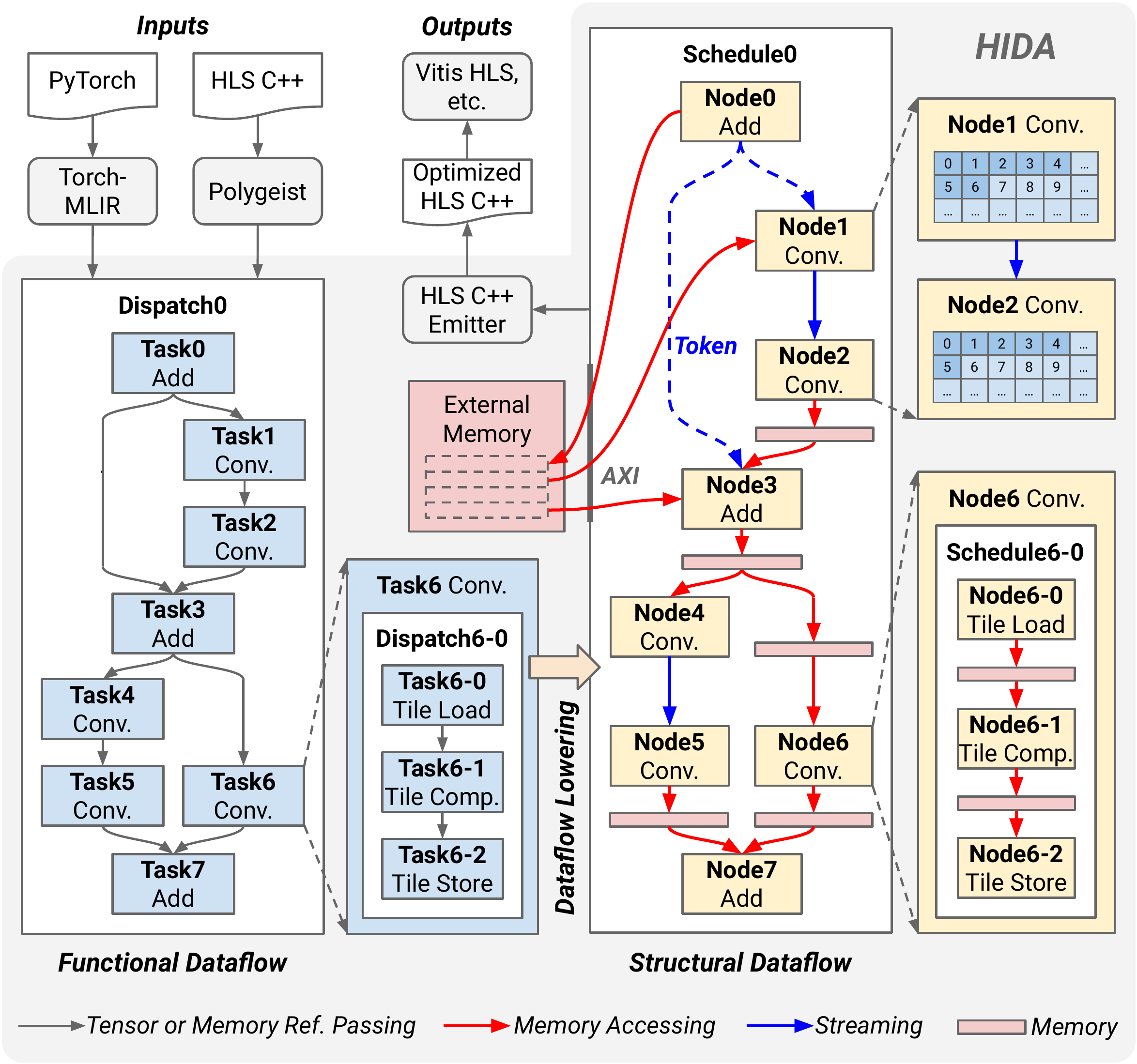}
    \vspace{-16pt}
    \caption{HIDA framework overview.}
    \label{fig:framework}
\end{minipage}
\hfill
\begin{minipage}{0.43\textwidth}
    \centering
    \small
    \captionof{table}{HIDA-IR key operations. \emph{Region} is a sequential list of operations to be executed.}
    \label{tab:dialect}
    \vspace{-4pt}
    \setlength\tabcolsep{2pt}
    \begin{tabular}{cp{18em}}
        \toprule
        \textbf{Operation} & \multicolumn{1}{c}{\textbf{Description}} \\
        \midrule
        \multicolumn{2}{c}{\textbf{Functional Dataflow}} \\
        \midrule
        \texttt{task} & Own a transparent region, can contain nested dispatch operations with sub-tasks. \\
        \hline
        \texttt{dispatch} & Launch multiple tasks in its region. \\
        \midrule
        \multicolumn{2}{c}{\textbf{Structural Dataflow}} \\
        \midrule
        \texttt{node} & Own an isolated region, can contain nested schedule operations with sub-nodes. Carry explicit I/O memory effect information. \\
        \hline
        \texttt{schedule} & An isolated region with multiple nodes. Carry explicit scheduling information. \\
        \hline
        \texttt{buffer} & A buffer with variadic stages and ports and automatic ping-pong buffering semantics. Carry explicit partition and layout information. \\
        \hline
        \texttt{stream} & A stream channel with variadic entries. \\
        \midrule
        \multicolumn{2}{c}{\textbf{Module Interface}} \\
        \midrule
        \texttt{port} & A memory or stream port with explicit type. \\
        \hline
        \texttt{bundle} & A named bundle of ports. \\
        \hline
        \texttt{pack} & Pack an external memory block into a port. \\
        \bottomrule
    \end{tabular}
\end{minipage}
\end{figure*}

\section{Background}
\label{sec:background}

\subsection{MLIR Framework}
MLIR~\cite{lattner2020mlir, mlir2023github} is a compilation framework supporting multiple levels of functional and representational hierarchy. In the remainder of this paper, we use \emph{MLIR} to refer to the MLIR framework and \emph{IR} for the intermediate representation of programs in MLIR. MLIR includes a single static assignment (SSA) style IR~\cite{cytron1991efficiently} where an \emph{Operation} is the minimal unit of code. Each operation accepts a set of typed \emph{Operand}s and produces a set of typed \emph{Result}s. Connections between the results of one operation and the operands of another operation describe the SSA-style flow of data. For instance, \texttt{\%3 = addi \%0, \%c1} in Figure \ref{fig:mlir}(iii) is an operation with operands \texttt{\%0} and \texttt{\%c1} and result \texttt{\%3}. Each operation can also be parameterized by a set of \emph{Attribute}s indicating important characteristics of the operation. Unlike operands, which typically model values produced by other operations when a program is executed, attributes have values that are known and fixed at compile time. A sequential list of operations without control flow is defined as a \emph{Block} and a control flow graph (CFG) of blocks is organized into a \emph{Region} in MLIR. Regions are, in turn, contained by operations, enabling the description of arbitrary design hierarchy. In MLIR, \emph{Function} is defined as a built-in callable operation always owning one region. For instance, function \texttt{@foo} in Figure \ref{fig:mlir}(iii) owns one region containing four blocks, \texttt{bb0} to \texttt{bb3}.

A \emph{Dialect} in MLIR defines a namespace for a group of related operations, attributes, and types. MLIR not only provides multiple built-in dialects to represent common functionalities, but also features an open infrastructure allowing to define new dialects at different abstraction levels. \emph{Pass} is a key component of compiler which traverses the IR for the purpose of optimization or analysis. Similar to LLVM~\cite{lattner2004llvm}, users can design \emph{Transform} and \emph{Analysis} passes in MLIR to perform the IR transformation and analysis. However, in the context of MLIR, \emph{Transform} typically refers to the transformation within a dialect. The transformation between different dialects is typically referred as \emph{Conversion}, while the transformation between MLIR and external representation is referred as \emph{Translation}. \emph{Lowering} is a terminology referring to the process of lowering the abstraction level of IR.

\subsection{Relevant MLIR Dialects}
\label{subsec:mlir_dialects}
Many dialects in MLIR are immediately applicable for representing nested loop programs commonly used in HLS. The \texttt{linalg} dialect provides a structured representation of linear algebra operations. The \texttt{affine} dialect provides a powerful abstraction for affine operations in order to make dependence analysis and loop transformations efficient and reliable. The \texttt{affine} dialect defines \emph{Affine Map} as a mathematical function that transforms a list of affine values into a list of results. Affine operations (e.g., \texttt{affine.for} and \texttt{if}) must take affine values as input operands, therefore the loop bounds of \texttt{affine.for} operation and conditions of \texttt{affine.if} operation must be the expression of affine values. The \texttt{scf} (structured control flow) dialect defines control flow operations (e.g., \texttt{scf.for} and \texttt{if}) whose loop bounds or conditions can be any SSA values. Therefore, \texttt{scf} operations are not constrained by the affine requirements and can represent a wider range of programs. MLIR also provides several fundamental built-in dialects to represent basic arithmetic operations (e.g., \texttt{addf}), unstructured control flow operations (e.g., \texttt{br} and \texttt{cond\_br}), and memory-related operations (e.g., \texttt{load} and \texttt{store}). Taking Figure \ref{fig:mlir} as an example, the structured control flows in Figure \ref{fig:mlir}(i) and (ii) represented with \texttt{affine} and \texttt{scf} operations are flattened to the unstructured \texttt{br} and \texttt{cond\_br} operations in Figure \ref{fig:mlir}(iii).

\begin{figure*}
    \centering
    \includegraphics[width=\textwidth]{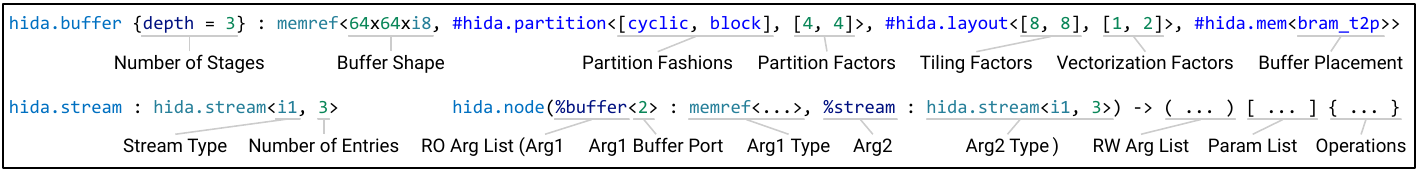}
    \vspace{-16pt}
    \caption{\texttt{buffer}, \texttt{stream}, and \texttt{node} operation syntax in \emph{Structural} dataflow. \emph{RO} and \emph{RW} denote read-only and read-write.}
    \label{fig:syntax}
\end{figure*}

\section{HIDA Overview}
\label{sec:overview}

Figure \ref{fig:framework} shows the overall architecture of HIDA. HIDA is built on top of the MLIR infrastructure~\cite{lattner2021mlir, mlir2023github} and can take deep learning models written in PyTorch~\cite{paszke2019pytorch} or generic HLS C++ code as design entries and produce optimized HLS C++ code. For the PyTorch and C++ inputs, we use Torch-MLIR~\cite{torchmlir2023github} and Polygeist~\cite{moses2021polygeist} as front-ends to parse source codes. After the optimizations are completed in HIDA, we use an HLS C++ emitter~\cite{hpca2022scalehls} to generate synthesizable HLS C++ code, which can then be mapped to RTL designs with downstream HLS tools~\cite{vitishls2022userguide, intelhls2022userguide, legup2021document}. HIDA proposes two new techniques to handle the \emph{representation} and \emph{optimization} of dataflow compilation, which are the key enablers to tackle the challenges discussed in Section~\ref{sec:motivation}:
\begin{itemize}[leftmargin=*]
    \item \emph{Hierarchical Dataflow IR (HIDA-IR).} As shown in Figure~\ref{fig:framework}, HIDA consists of \emph{Functional} and \emph{Structural} dataflow IR carved for different purposes. Table~\ref{tab:dialect} summarizes the key operations of HIDA-IR. Details can be found in Section~\ref{sec:ir}.
    \item \emph{Hierarchical Dataflow Optimizer (HIDA-OPT).} HIDA decouples the HLS optimization problems of \emph{Functional} and \emph{Structural} dataflow to handle HLS designs at scale. Details can be found in Section~\ref{sec:opt}.
\end{itemize}

\section{HIDA-IR}
\label{sec:ir}

Currently, HLS tools~\cite{zhao2017comba, hpca2022scalehls, zhao2022polsca} employ call graphs to represent HLS structures using sequential IRs. However, due to the lack of expressiveness for the parallel characteristics and micro-architecture of dataflow, these IRs have limited capability when used in scalable dataflow optimization flows. To address this problem, we propose a holistic HIDA-IR with two levels of representation, which we refer to as \emph{Functional} and \emph{Structural} dataflow. The \emph{Functional} dataflow is designed to capture the high-level characteristics and hierarchy of HLS designs, driving the algorithmic optimizations and task fusion. In contrast, the \emph{Structural} dataflow is a low-level abstraction that captures the micro-architectural details and is optimized to handle the scheduling and parallelization.

\subsection{Functional Dataflow}
\textbf{Hierarchical Structure.}
In the \emph{Functional} dataflow, we introduce a \texttt{dispatch} operation that contains the computation graph to be dispatched. Within the \texttt{dispatch} operation, all graph nodes are partitioned into multiple \texttt{task} operations to represent the dispatch strategy. As accelerators often have multiple levels of dataflow to achieve higher parallelism, HIDA-IR supports a hierarchical structure by allowing the recursive nesting of \texttt{task} and \texttt{dispatch} operations. Figure~\ref{fig:framework} visualizes a hierarchical \emph{Functional} dataflow, where \emph{Task6} contains three sub-tasks that can be executed in a dataflow manner to hide the latency of loads and stores, allowing us to utilize the computational capability to a greater extent.

\textbf{Transparent from Above.}
At the \emph{Functional} level, tasks often need to be manipulated as different dispatch strategies can lead to different trade-offs. Based on this observation, the \texttt{dispatch} and \texttt{task} operations are designed to be transparent and share the global context, simplifying the process of the fusing and splitting of tasks. As a result, we can efficiently explore various dispatch strategies at the \emph{Functional} level. Meanwhile, thanks to the transparency, buffers and tensors defined in the global context can be accessed by tasks at all hierarchies without indirection. Therefore, the \emph{Functional} dataflow can enable effective algorithmic optimizations for both PyTorch and C++ programs, which model the computations at tensor and memory levels, respectively.

\subsection{Structural Dataflow}
\textbf{Memory-Mapped and Stream Buffer.}
In order to precisely capture the on-chip and off-chip memory accessing behaviors, we introduce two types of buffers at the \emph{Structural} dataflow level, the memory-mapped buffer and the stream buffer, which are represented with the \texttt{buffer} and \texttt{stream} operations, respectively. Figure~\ref{fig:syntax} shows their syntax, where \texttt{\%} denotes a single static assignment~(SSA) value~\cite{cytron1991efficiently}. The embedded partition and data layout attributes of the \texttt{buffer} operation are designed to be converted to semi-affine maps~\cite{affine2023document}, enabling polyhedral-based dependency analysis and transformation~\cite{benabderrahmane2010polyhedral} in HIDA-OPT. Notably, to facilitate dataflow optimizations, \texttt{buffer} operations inherently carry ping-pong buffering semantics, allowing buffer instances to interleave the accesses from producers and consumers to improve the communication efficiency. Figure~\ref{fig:framework} visualizes the combined usage of the two types of buffers, where the red boxes and blue arrows represent the memory-mapped buffer and stream buffers. Dashed blue arrows denote single-bit stream buffers. In addition to the buffers, we introduce the \texttt{port} operation to represent memory-mapped or stream interfaces, capturing the interface characteristics, such as latency, that can have a considerable impact on the dataflow efficiency. For instance, in Figure~\ref{fig:framework}, \emph{Node0}, \emph{Node1}, and \emph{Node2} are scheduled to communicate through external memory, where the AXI interfaces are modeled with \texttt{port} operations.

\textbf{Isolated from Above.}
In the \emph{Structural} dataflow, we introduce \texttt{schedule} and \texttt{node} operation as the counterparts of \texttt{dispatch} and \texttt{task} operation. While \emph{Structural} dataflow has a hierarchical structure similar to \emph{Functional} dataflow, an important distinction between the two is that the \texttt{schedule} and \texttt{node} operations are isolated from the external context. Therefore, external values must be passed into \texttt{schedule} and \texttt{node} as arguments. In addition, the \texttt{node} operation carries explicit memory effect information for each argument to avoid unnecessary inter-node effect analysis. Figure~\ref{fig:syntax} shows the syntax of \texttt{node} operation, where \texttt{\%buffer} and \texttt{\%stream} are passed in as read-only arguments in this specific example. For simplicity, we omit the read-write arguments and constant parameters in Figure~\ref{fig:syntax}. The rationale behind this design decision is driven by how the \emph{Structural} dataflow carries the architecture optimization. By isolating the context of \texttt{schedule} and \texttt{node}, the dataflow optimization problem can be cleanly partitioned into multiple local intra-node optimizations and global inter-node optimization, leading to a decoupled scalable solution. Details of the optimization process is elaborated in Section~\ref{sec:opt}.

\begin{figure}
    \centering
    \includegraphics[width=\linewidth]{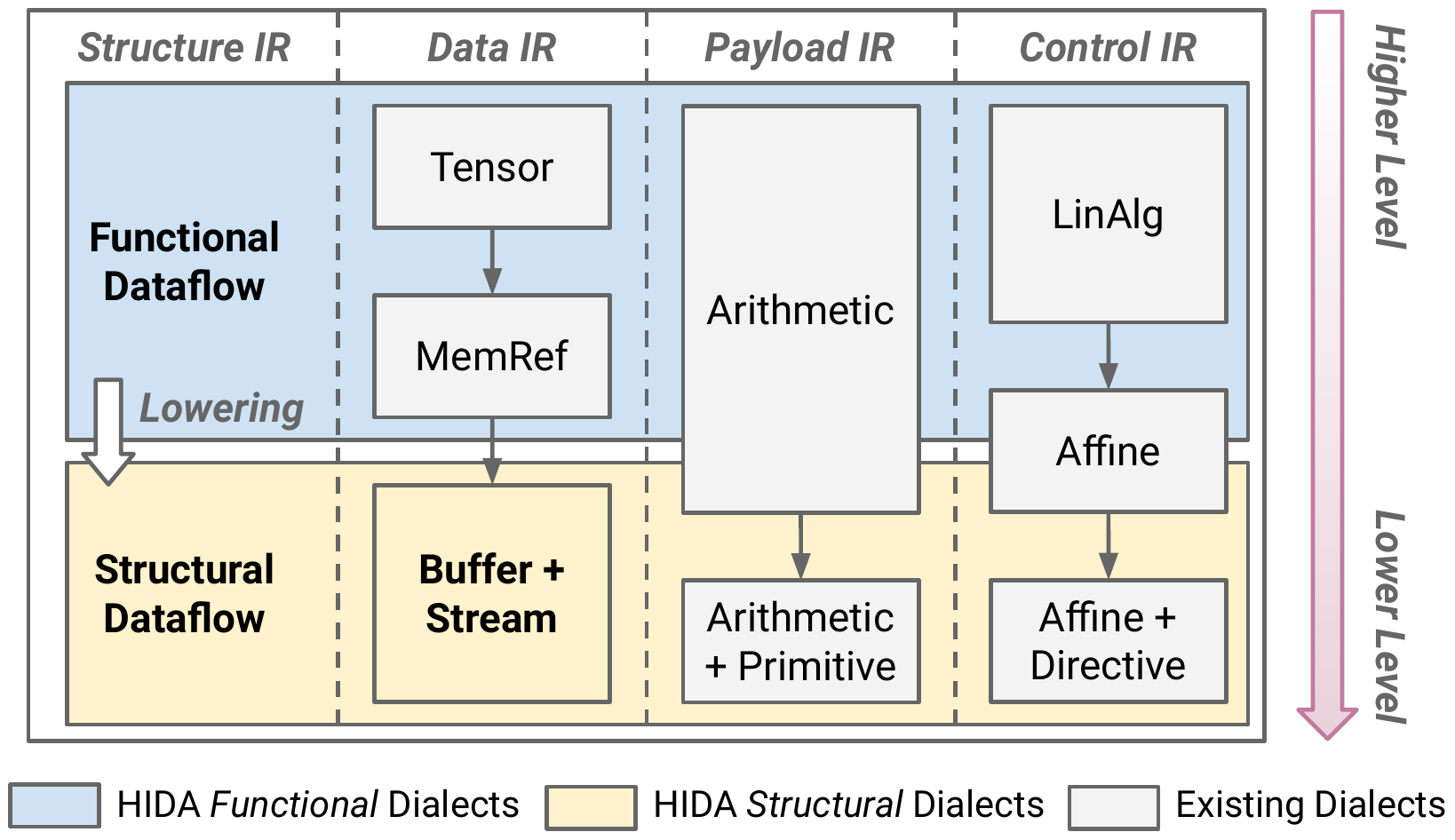}
    \vspace{-17pt}
    \caption{Integration with MLIR dialects.}
    \label{fig:dialects}
\end{figure}

\subsection{Integration with MLIR Dialects}
HIDA reuses a set of MLIR built-in dialects to represent the common program components in PyTorch or C/C++. Figure~\ref{fig:dialects} illustrates the integration of HIDA with the built-in MLIR dialects. The vertical axis of Figure~\ref{fig:dialects} represents the various abstraction levels of the dialects. Horizontally, all dialects are categorized into four different types: \emph{structure}, \emph{data}, \emph{payload}, and \emph{control-flow}. At every abstraction level, the corresponding four types of dialects are combined with each other to represent the complete functionality of a program. The \emph{Functional} dataflow is first combined with \texttt{tensor}, \texttt{arith}, and \texttt{linalg} dialects to represent the tensor-level programs compiled from PyTorch. Then, the \texttt{tensor} and \texttt{linalg} dialects are lowered to \texttt{memref} and \texttt{affine} dialects, respectively, while the lowered IRs are still combined with the \emph{Functional} dataflow due to its adaptability with both tensor and memory semantics. Once the optimizations at the \emph{Functional} dataflow level are completed, the \emph{Functional} dataflow is lowered to the \emph{Structural} dataflow, while the \texttt{memref} operations are lowered to the \emph{Structural} buffer and stream operations. Note that the \texttt{affine} dialect is used in both the \emph{Functional} and \emph{Structural} dataflows for loop analyses and transformations. HIDA reuses the \emph{Primitive} and \emph{Directive} IRs from ScaleHLS~\cite{hpca2022scalehls} to represent the HLS-specific structures, such as the loop pipelining directive. By integrating with the existing dialects, HIDA can carry out loop and directive-level HLS optimizations in a hierarchical manner.

\section{HIDA-OPT}
\label{sec:opt}

Although previous works~\cite{hpca2022scalehls, ejjeh2022hpvm2fpga} have enabled inter-task parallelization through dataflow, they could not conduct dataflow-oriented optimizations. Specifically, ScaleHLS~\cite{hpca2022scalehls} could automatically legalize a computation graph into a dataflow model and enable code generation but ignored the inter-task design space coupling~\cite{jun2023autoscaledse}, resulting in suboptimal dataflow designs. HPVM2FPGA~\cite{ejjeh2022hpvm2fpga} used an ML-driven algorithm~\cite{nardi2019practical} for DSE, but it only introduced a boolean parameter to enable/disable global dataflow, again, leading to the restricted search of the design space.
In this section, we propose a HIDA-OPT solution consisting of five steps to tackle the dataflow optimization problem.

\begin{algorithm}[t]
\small
\captionsetup{font=small}
\caption{Functional dataflow construction}
\label{alg:dataflow_construction}
\begin{algorithmic}[1]
\Require $m$, top module of the initial computation graph
\Ensure Updated $m$, top module of the \emph{Functional} dataflow

\For{$n$ \textbf{in} postorder\_walk($m$, has\_region())}
\If{is\_dispatchable(get\_region($n$))}
\State $d \gets$ wrap\_ops(get\_ops($n$), new(\texttt{dispatch}))
\For{$op$ \textbf{in} get\_ops($d$)}
\State wrap\_ops(\{$op$\}, new(\texttt{task}))
\EndFor
\EndIf
\EndFor
\end{algorithmic}
\end{algorithm}

\subsection{Functional Dataflow Construction}
We leverage the transformations available in MLIR~\cite{lattner2021mlir, mlir2023github} to generate a \emph{hierarchical} computation graph at the level of linear algebra~\cite{linalg2023document} or loop~\cite{affine2023document}. Algorithm~\ref{alg:dataflow_construction} shows the pseudo-code of converting the initial computation graph to the \emph{Functional} dataflow. From line~1 to 3, we wrap each \emph{dispatchable} region with a \texttt{dispatch} operation in a bottom-up manner, where a region is defined as \emph{dispatchable} if it is owned by an iterative operation, such as \texttt{loop} and \texttt{func}, while containing at least two iterative operations. For instance, a loop region containing two child loops is considered to be \emph{dispatchable} as the two child loops can be dataflowed. Then, from line~4 to 5, each operation is wrapped with a separate \texttt{task} operation to construct a legal dataflow model.

\begin{algorithm}[t]
\small
\captionsetup{font=small}
\caption{Functional dataflow task fusion}
\label{alg:task_partition}
\begin{algorithmic}[1]
\Require $m$, top module of the \emph{Functional} dataflow
\Require $patterns$, set of profitable task fusion patterns
\Ensure Updated $m$, top module of the partitioned dataflow

\For{$d$ \textbf{in} preorder\_walk($m$, is\_instance(\texttt{dispatch}))}
\State $worklist \gets$ queue(get\_tasks($d$))
\While{\textbf{not} is\_empty($worklist$)}
\State $t \gets$ pop($worklist$)
\If{$t' \gets$ get\_matched\_task($t$, $patterns$)}
\State push($worklist$, wrap\_ops(\{$t$, $t'$\}, new(\texttt{task})))
\EndIf
\EndWhile
\Repeat{~$t_0, t_1 \gets$ get\_least\_critical\_tasks($d$, 2)}
\State wrap\_ops(\{$t_0$, $t_1$\}, new(\texttt{task}))
\Until{\textbf{not} is\_fusion\_profitable($t_0, t_1$)}
\State simplify\_dispatch\_hierarchy($d$)
\EndFor
\end{algorithmic}
\end{algorithm}

\subsection{Functional Dataflow Optimization}
Once the initial \emph{Functional} dataflow is constructed, we can optionally fuse dataflow tasks to balance the task workloads while reducing the communication cost. Algorithm~\ref{alg:task_partition} shows the task fusion process, where the inputs are the initial dataflow and a set of pre-defined profitable fusion patterns, such as element-wise operations fusion. From line~1 to 10, we recursively partition each \texttt{dispatch} operation in a top-down manner. Specifically, we first fuse adjacent tasks into new tasks through a pattern-driven worklist algorithm (lines~2 to 6). The pre-defined task fusion patterns are recursively applied to the dataflow until no pattern can be matched. Then, we continuously fuse the least critical two adjacent tasks to re-balance the dataflow until the fusion begins to generate a new critical task (lines~7 to 9). Finally, in line~10, we simplify the hierarchy by canonicalizing the \texttt{dispatch} and \texttt{task} operations. For instance, a task containing only one sub-task should be canonicalized to a single task. Notably, HIDA-IR’s systematic dataflow representation allows the task fusion process to be expanded with different heuristics or algorithms under varying scenarios.

\begin{figure}[t]
    \centering
    \includegraphics[width=0.9\linewidth]{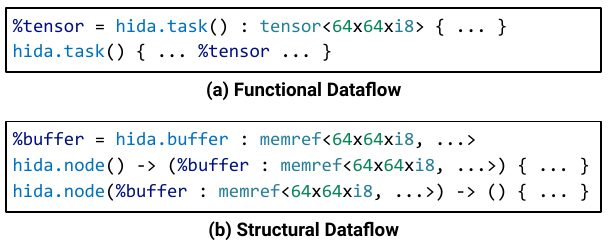}
    \vspace{-8pt}
    \caption{\emph{Functional} to \emph{Structural} dataflow lowering.}
    \label{fig:lowering}
\end{figure}

\subsection{Structural Dataflow Construction}
The \emph{Functional} dataflow can be lowered to \emph{Structural} dataflow for low-level optimizations and code generation. The dataflow lowering is composed of three procedures: (1)~\texttt{buffer} operation generation; (2)~\texttt{dispatch} to \texttt{schedule} operation mapping; (3)~\texttt{task} to \texttt{node} operation mapping. Figure~\ref{fig:lowering} shows a simplified example of the \emph{Functional} to \emph{Structural} dataflow lowering, where the two \texttt{task} operations in Figure~\ref{fig:lowering}(a) are mapped to \texttt{node} operations in Figure~\ref{fig:lowering}(b) correspondingly. The distinction between tensor and memory semantics draws a line between the \emph{Functional} and \emph{Structural} dataflow: tensors are immutable objects passed between producers and consumers directly; buffers are mutable objects that can be instantiated in hardware and modified multiple times. Therefore, in Figure~\ref{fig:lowering}(a), the \texttt{\%tensor} object is produced by the first \texttt{task} operation and directly consumed inside of the second \texttt{task} operation. In contrast, we can observe that the original \texttt{\%tensor} object is lowered to a \texttt{buffer} operation in Figure~\ref{fig:lowering}(b). Correspondingly, the first \texttt{node} operation becomes a user of the \texttt{\%buffer} object with read-write effect, while the second \texttt{node} operation uses the \texttt{\%buffer} object with read-only effect, given the syntax defined in Figure~\ref{fig:syntax}.

\textbf{Automation.}
For procedure (1) above, each tensor result of each \texttt{task} operation is converted to a \texttt{buffer} operation with default partition fashions, tiling, vectorization factors, and placement annotations. For procedures (2) and (3) above, because the \emph{Functional} operations are transparent while the \emph{Structural} operations are isolated from above context, the live-ins and memory effects are analyzed during the mapping. As shown in Figure~\ref{fig:syntax}, the arguments of the \texttt{node} operations are explicitly grouped based on their memory effects. Notably, procedure (1) to (3) can generate a legal \emph{Structural} dataflow, and the \texttt{node} and \texttt{buffer} operations will be optimized later in the dataflow balancing and parallelization.

\begin{figure}
    \centering
    \includegraphics[width=0.9\linewidth]{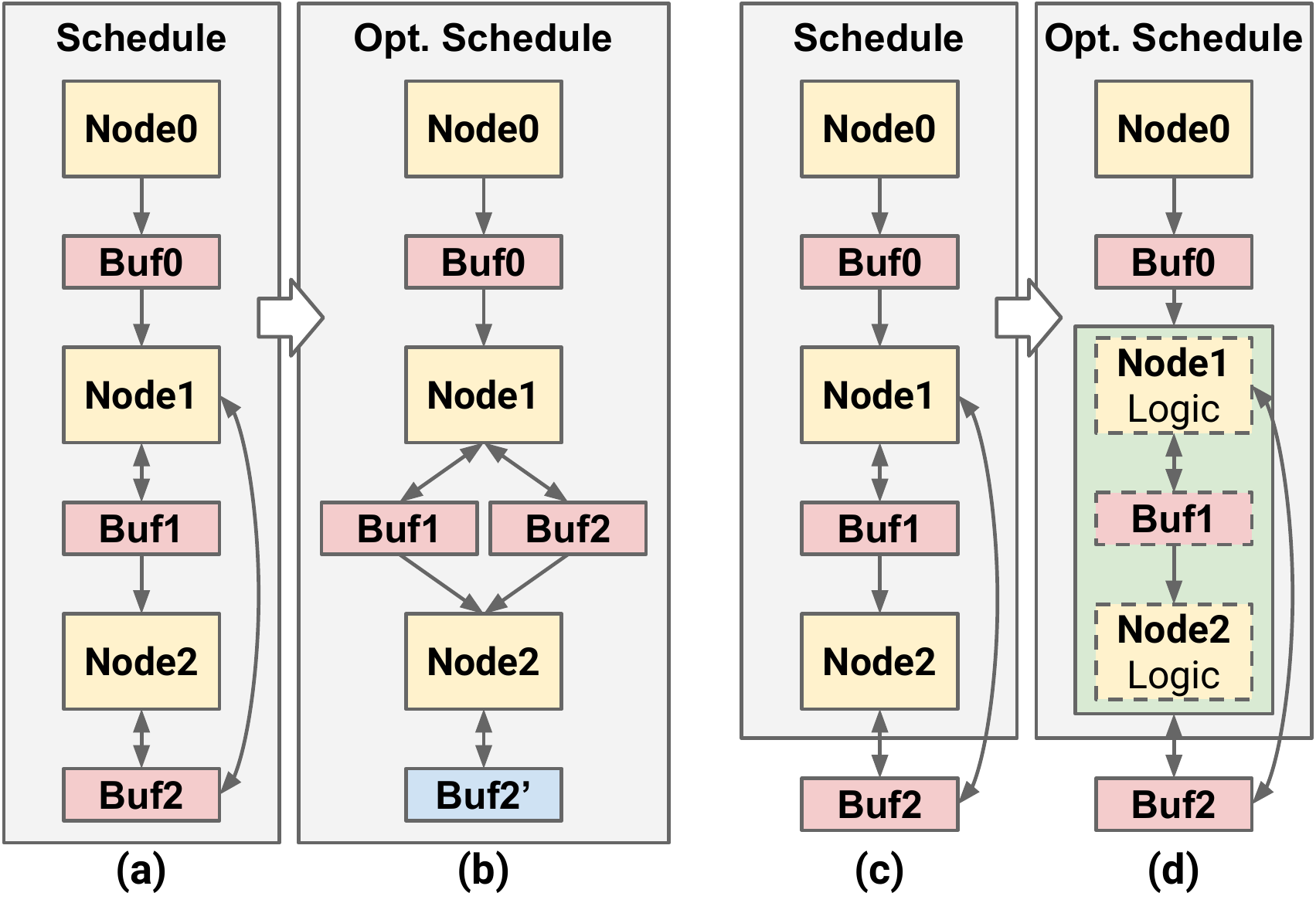}
    \vspace{-6pt}
    \caption{Eliminate multiple producers. Double arrows indicate both read and write on a buffer. For instance, in (a), \emph{Node1} may first read from \emph{Buf2} and then write results to it.}
    \label{fig:multi_producer}
\end{figure}

\subsection{Structural Dataflow Optimization}
At the \emph{Structural} dataflow level, we propose two optimizations that are crucial for dataflow efficiency but have not been thoroughly studied in existing tools: \emph{(1)~Multi-producer elimination}, which can eliminate multiple nodes writing to the same buffer and improve the dataflow parallelism; \emph{(2)~Data path balancing}, which can balance different data paths by inserting on-chip or external buffers, balancing the pipeline execution rate to achieve the best throughput.

\subsubsection{Eliminate Multiple Producers}
Dataflow architectures often contain buffers written to by multiple producers, leading to inefficient dataflow execution. For example, in Figure~\ref{fig:multi_producer}(a), correctly managing the memory access of \emph{Buf2} is challenging as \emph{Node1} and \emph{Node2} simultaneously write to it. As a result, to preserve correctness, the dataflow structure must be executed sequentially.

\begin{algorithm}[t]
\small
\captionsetup{font=small}
\caption{Multiple producers elimination}
\label{alg:multi_producer}
\begin{algorithmic}[1]
\Require $s$, dataflow schedule
\Ensure Updated $s$, transformed dataflow schedule

\For{$b$ \textbf{in} get\_internal\_buffers($s$)}
\State $p\_list \gets$ topo\_sort(get\_producers($b$))
\For{$p$ \textbf{in} drop\_front($p\_list$)} \Comment{exclude the 1st producer}
\State $b' \gets$ clone($b$)
\If{read\_effect($p$, $b$)}
\State $copy \gets$ new(\texttt{copy}, $b$, $b'$)
\State insert\_to\_front($copy$, get\_region($p$))
\EndIf
\For{$u$ \textbf{in} get\_users($b$)}
\If{dominate($p$, $u$)} \Comment{include $p$ itself}
\State replace\_use\_with($b$, $b'$, $u$)
\EndIf
\EndFor
\EndFor
\EndFor

\For{$b$ \textbf{in} get\_external\_buffers($s$)}
\State $p\_list \gets$ get\_producers($b$)
\State wrap\_ops($p\_list$, new(\texttt{node})) \Comment{merge producers}
\EndFor
\end{algorithmic}
\end{algorithm}

\textbf{Solution.}
HIDA-OPT resolves this problem considering two cases: \emph{(1)~Buffer duplication.} In the case of Figure~\ref{fig:multi_producer}(a), we eliminate the multiple producers by duplicating \emph{Buf2} into the \emph{Buf2'} as shown in Figure~\ref{fig:multi_producer}(b). As a result, \emph{Node1} and \emph{Node2} no longer write to \emph{Buf2} simultaneously, allowing the structure to be scheduled in a pipelined manner. This duplication is possible due to the semantics of the \emph{Structural} IR, where \emph{Buf2} is allocated inside the context of its parent schedule, ensuring that no external side-effect operation can access \emph{Buf2}. \emph{(2)~Node fusion.} For Figure~\ref{fig:multi_producer}(c), \emph{Node1} and \emph{Node2} write to \emph{Buf2} allocated outside of its parent schedule. In this case, we cannot apply the same transformation as case~(1) because there may exist an external node having write-effect on \emph{Buf2}. Specifically, if we duplicate \emph{Buf2} into \emph{Buf2'}, only the original \emph{Buf2} can be updated by the external write-effect nodes, leaving outdated data in \emph{Buf2'}. Therefore, in the next iteration, \emph{Buf2'} may hold incorrect data and eventually lead to incorrect functionality. Thus, to eliminate the multi-producer violation, we fuse \emph{Node1} and \emph{Node2} into a new node and sequentially execute them as shown in Figure~\ref{fig:multi_producer}(d).

\textbf{Automation.}
Lines 1 to 10 of Algorithm~\ref{alg:multi_producer} show the pseudo-code of case~(1) above. For each internal buffer, we first collect all its \emph{producer}s and sort them based on SSA dominance to maintain the predetermined memory access order in the subsequent transformations. Then, for each producer except the first one, we duplicate a new buffer \emph{b'} for it (line 4) and create an explicit memory copy from the original buffer \emph{b} to \emph{b'} (lines 5 to 7) if the current producer \emph{p} reads from \emph{b}. Finally, we replace uses of \emph{b} with \emph{b'} if the user is dominated by the \emph{p} (lines 8 to 10), leaving exactly one producer for the original buffer \emph{b}. Lines 11 to 13 of Algorithm~\ref{alg:multi_producer} shows the pseudo-code of case~(2) that merges all producers of each external buffer into a single \texttt{node} to avoid data racing.

\textbf{Complexity.}
One can observe that case~(2) employs a more conservative transformation and enables the dataflow execution of the whole design by only enforcing the sequential execution of \emph{Node1} and \emph{Node2}. One may argue for a more comprehensive inter-node analysis of \emph{Buf2} to determine whether the external write-effect nodes will intervene if we were to duplicate \emph{Buf2} - this could work on small dataflow architectures, but it does not scale well. Many dataflow nodes at different hierarchies can access shared buffers; as a result, an inter-node analysis has a complexity of $O(mn^2)$, where $m$ denotes the number of shared buffers and $n$ is the number of nodes accessing the same buffer.

\begin{figure}
    \centering
    \includegraphics[width=0.95\linewidth]{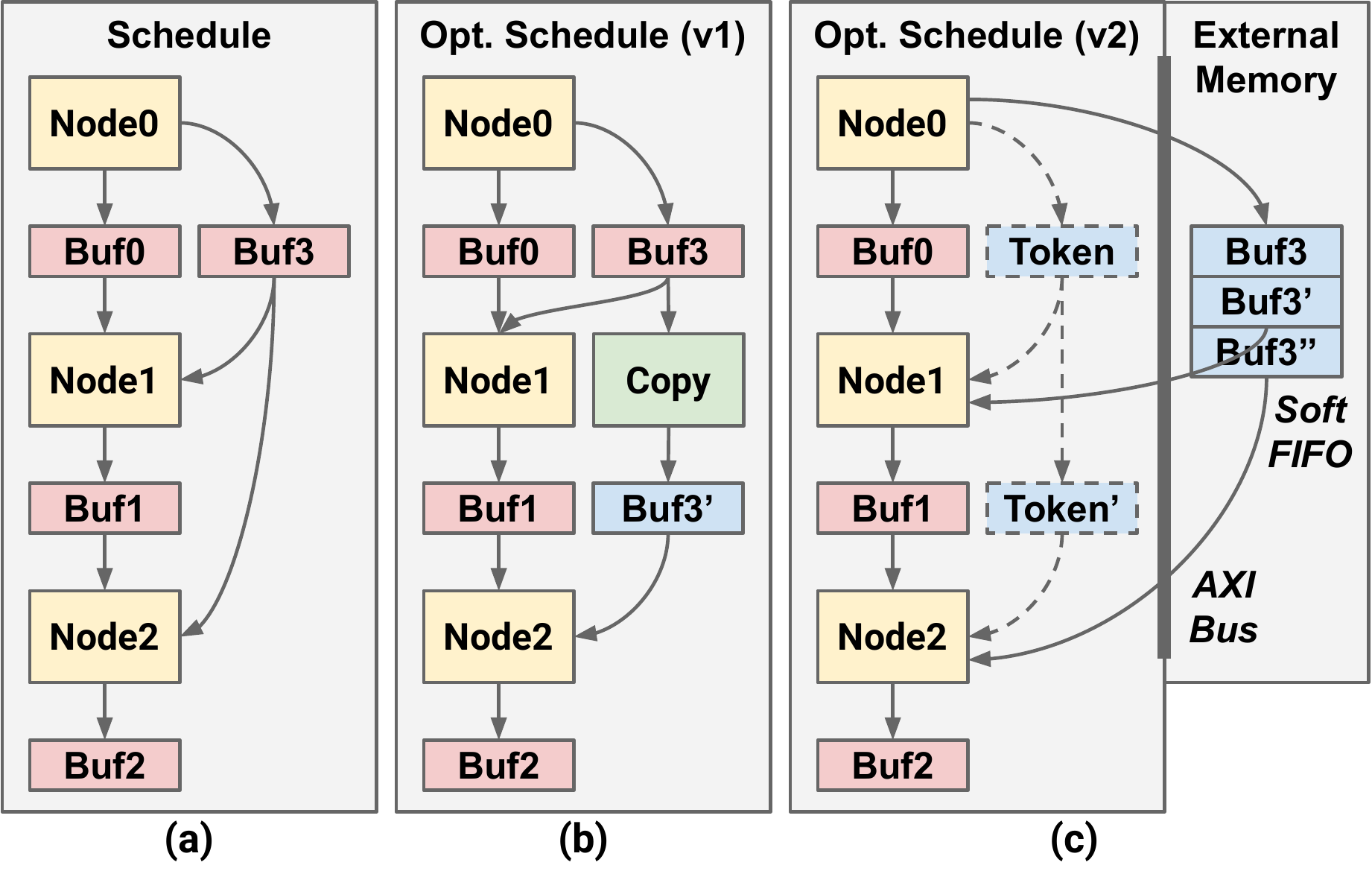}
    \vspace{-6pt}
    \caption{Balance data paths. Dash line block represents a 1-bit token buffer. Soft FIFO is allocated in external memory and interfaced with dataflow through AXI interconnect.}
    \label{fig:balance_dataflow}
\end{figure}

\subsubsection{Balance Data Path}
A complicated dataflow structure often has multiple data paths; some paths may have more dataflow nodes to execute. If left unoptimized, the unbalanced paths can significantly degrade the overall performance of the final design. For instance, in Figure~\ref{fig:balance_dataflow}(a), the \emph{Node0-Node1-Node2} path is longer than the \emph{Node0-Node2} path. As a result, \emph{Node0} must wait until the longer path completes before it can process the next data frame. This situation is very common in real-world applications, such as ResNet~\cite{he2016deep}, which has shortcut paths in the residual blocks. Note that there are two levels of balancing in HIDA: one is the \emph{data path} balancing we discuss in this section; the other is \emph{node delay} balancing that will be handled separately during the dataflow parallelization. 

\textbf{Solution.}
HIDA-OPT resolves this issue using two methods: \emph{(1)~On-chip buffer duplication.} We can duplicate buffers on the shorter data paths to balance the execution speed. For instance, in Figure~\ref{fig:balance_dataflow}(b), \emph{Buf3} is duplicated to \emph{Buf3'}, followed by an automatic insertion of a copy node between \emph{Buf3} and \emph{Buf3'}. Through this approach, the \emph{Node0-Copy-Node2} data path can execute in a pipelined manner at the same rate as another data path, such that \emph{Node0} no longer needs to wait. \emph{(2)~Soft FIFO in external memory.} As shown in Figure~\ref{fig:balance_dataflow}(c), a soft FIFO is allocated in the external memory to substitute \emph{Buf3}. The FIFO is \emph{soft} because data is not really shifted in the FIFO. Instead, the memory access addresses of dataflow nodes are rotated to access the correct data. For instance, in Figure~\ref{fig:balance_dataflow}(c), \emph{Node0} is writing to \emph{Buf3}, while \emph{Node1} and \emph{Node2} are reading from \emph{Buf3'} and \emph{Buf3''}, respectively. Then in the next dataflow iteration, \emph{Node0} will write to \emph{Buf3''}, and \emph{Node1} and \emph{Node2} will read from \emph{Buf3} and \emph{Buf3'}, respectively.

\begin{table*}
\begin{minipage}{0.4\textwidth}
\captionof{lstlisting}{A dataflow example in C++. We assume each nested loop is a dataflow node.}
\begin{lstlisting} [language=C++]
float A[32][16];
NODE0_I: for (int i=0; i<32; i++)
  NODE0_K: for (int k=0; k<16; k++)
    A[i][k] = ...; // Load array A.

float B[16][16];
NODE1_K: for (int k=0; k<16; k++)
  NODE1_J: for (int j=0; j<16; j++)
    B[k][j] = ...; // Load array B.

float C[16][16];
NODE2_I: for (int i=0; i<16; i++)
  NODE2_J: for (int j=0; j<16; j++)
    NODE2_K: for (int k=0; k<16; k++)
      C[i][j] = A[i*2][k] * B[k][j];
\end{lstlisting}
\label{code:dataflow_cpp}
\end{minipage}
\hfill
\begin{minipage}{0.58\textwidth}
\centering
\small
\caption{Node connections appearing in Listing~\ref{code:dataflow_cpp}. \emph{S} and \emph{T} denote the source and target nodes of the connection. $\emptyset$ denotes empty.}
\label{tab:connections}
\vspace{-4pt}
\setlength\tabcolsep{4pt}
\begin{tabular}{ccccccc}
    \toprule
    \multirow{2}{*}[-2pt]{\textbf{Source}} &
    \multirow{2}{*}[-2pt]{\textbf{Target}} &
     \multirow{2}{*}[-2pt]{\textbf{Buffer}} &
    \multicolumn{2}{c}{\textbf{Permutation Map}} &
    \multicolumn{2}{c}{\textbf{Scaling Map}} \\
    \cmidrule(lr){4-5} \cmidrule(lr){6-7}
    & & & \textbf{S-to-T} & \textbf{T-to-S} & \textbf{S-to-T} & \textbf{T-to-S} \\
    \midrule
    Node0 & Node2 & A & [0, $\emptyset$, 1] & [0, 2] & [0.5, 1] & [2, $\emptyset$, 1]\\
    Node1 & Node2 & B & [$\emptyset$, 1, 0] & [2, 1] & [1, 1] & [$\emptyset$, 1, 1]\\
    \bottomrule
\end{tabular}
\vspace{10pt}
\caption{Node parallelization results of Listing~\ref{code:dataflow_cpp} assuming a maximum parallel factor of 32.}
\label{tab:node_factors}
\vspace{-4pt}
\setlength\tabcolsep{3pt}
\begin{tabular}{cccccccc}
    \toprule
    \multirow{2}{*}[-2pt]{\textbf{Node}} &
    \multirow{2}{*}[-2pt]{\textbf{Intensity}} &
    \multicolumn{2}{c}{\textbf{Parallel Factor}} &
    \multicolumn{4}{c}{\textbf{Loop Unroll Factors}} \\
    \cmidrule(lr){3-4} \cmidrule(lr){5-8}
    & & \textbf{w/o IA} & \textbf{w/ IA} & \textbf{IA+CA} & \textbf{IA} & \textbf{CA} & \textbf{Naive} \\
    \midrule
    \textbf{Node0} & 512 & 32 & 4 & [4, 1] & [2, 2] & [8, 4] & [4, 8] \\
    \textbf{Node1} & 256 & 32 & 2 & [1, 2] & [1, 2] & [4, 8] & [4, 8] \\
    \textbf{Node2} & 4,096 & 32 & 32 & [4, 8, 1] & [4, 8, 1] & [4, 8, 1] & [4, 8, 1] \\
    \bottomrule
\end{tabular}
\end{minipage}
\end{table*}

\textbf{Elastic Node Execution.}
For method~(2) above, after the soft FIFO is generated, the original \emph{Buf3} is substituted with external memory interfaces, such as memory-mapped AXI interfaces. Therefore, the dependencies between dataflow nodes associated with \emph{Buf3} are no longer explicit - they access external memories through assigned addresses instead of sharing \emph{Buf3}. To maintain the correct execution order, HIDA can automatically construct a token flow between these dataflow nodes. For instance, in Figure~\ref{fig:balance_dataflow}(c), once \emph{Node0} completes its computation, it will send a \emph{Token}, and \emph{Node1} and \emph{Node2} will not start until they receive the \emph{Token} and \emph{Token'} respectively. This way, the token flows elastically maintain the execution order, and no static logic in the form of an FSM is needed to control the execution.

\subsection{Structural Dataflow Parallelization}
\label{sec:parallelization}
After dataflow optimization, HIDA will parallelize each node to improve the overall throughput and latency. However, automatic parallelization is often very challenging for several reasons: \emph{(1)~Memory data layout.} Degradation of dataflow performance may occur without proper alignment between the computation pattern and memory layouts. \emph{(2)~Connectedness of nodes.} We define two nodes as having a \emph{connection} if they communicate through shared buffers. Due to reason~(1), when two nodes are connected, the parallelism of each node should be aware of the shared memory layout. \emph{(3)~Computation intensity of nodes.} We define the number of operations contained by a node as its \emph{intensity}. To maximize the overall throughput while minimizing resource utilization, the optimal parallel factors should be proportional to the intensity of dataflow nodes. Due to the coupling of local design spaces, the local optimality of the scheduling of each node can no longer automatically lead to the global optimal solution for dataflow architectures.

\subsubsection{Intensity and Connection-Aware Approach}
The challenges discussed above are tightly coupled and need to be handled holistically. In HIDA, we propose an intensity-aware~(IA) and connection-aware~(CA) approach to determine the best parallelization strategies:

\textbf{Step~(1) Intensity and Connection Analysis.}
We first construct two maps to record the intensity and connections of each dataflow node. For each connection, we record the source and target node, associated buffer, permutation maps holding the loop level alignment, and scaling maps holding the stride alignment. For instance, in Listing~\ref{code:dataflow_cpp}, \emph{Node0} and \emph{Node2} are connected through array \emph{A}. Because \emph{Node0} writes to array \emph{A} with the first and second loops, while \emph{Node2} reads with the first and third loops, the \emph{Node0-to-Node2} permutation map is [0, $\emptyset$, 1], where $\emptyset$ denotes empty. Meanwhile, the \emph{Node0-to-Node2} scaling map is [0.5, 1] as \emph{Node2} reads array \emph{A} with a stride of 2. Table~\ref{tab:connections} shows the two connections in Listing~\ref{code:dataflow_cpp}. The permutation maps and scaling maps will be used to align the loop unroll factors of connected nodes in step~(4) below.

\textbf{Step~(2) Node Sorting.}
We then sort all dataflow nodes into a worklist in descending order of the number of connections with the computation intensity as tie-breaker. This determines the order in which we parallelize each node. Intuitively, the parallelization strategy of a node with more connections will affect more nodes, while higher-intensity nodes, being more computationally complex, are more sensitive to optimization. Therefore, following the results of step~(1), the order of nodes in Listing~\ref{code:dataflow_cpp} in terms of criticality is \emph{Node2}, \emph{Node0}, and then \emph{Node1}.

\textbf{Step~(3) Parallel Factor Generation.}
Guided by resource constraints, HIDA-OPT determines the maximum parallel factor that can be applied to a dataflow node. Then, the parallel factor of each node will be set proportionally to its intensity. In Listing~\ref{code:dataflow_cpp}, assuming the maximum parallel factor is 32, the parallel factor of each node is listed in Table~\ref{tab:node_factors}.

\begin{algorithm}[t]
\small
\captionsetup{font=small}
\caption{Node parallelization}
\label{alg:node_parallelization}
\begin{algorithmic}[1]
\Require $n$, dataflow node
\Ensure Updated $n$, transformed dataflow node

\State $constraints\_list \gets$ new([][]) \Comment{list of constraints}
\State $c\_list \gets$ get\_connected\_nodes($n$)
\For{$c$ \textbf{in} $c\_list$}
\State $unroll\_factors \gets$ get\_unroll\_factors($c$)
\State $s\_map \gets$ get\_scaling\_map($c$) \Comment{from step (1)}
\State $p\_map \gets$ get\_permutation\_map($c$) \Comment{from step (1)}
\State $constraints \gets$ permute($unroll\_factors \odot s\_map$, $p\_map$)
\State append($constraints\_list$, $constraints$)
\EndFor
\State $parallel\_factor \gets$ get\_parallel\_factor($n$) \Comment{from step (3)}
\State $dse \gets$ init\_dse($n$) \Comment{initialize DSE engine}
\Repeat{~$unroll\_factors \gets$ propose\_unroll\_factors($dse$)}
\State $is\_valid \gets$ \textbf{true}
\For{$constraints$ \textbf{in} $constraints\_list$}
\For{$constr$, $uf$ \textbf{in} zip($constraints$, $unroll\_factors$)}
\If{$constr$ \% $uf$ != 0 \textbf{and} $uf$ \% $constr$ != 0}
\State $is\_valid \gets$ \textbf{false}
\EndIf
\EndFor
\EndFor
\If{product($unroll\_factors$) >= $parallel\_factor$}
\State $is\_valid \gets$ \textbf{false}
\EndIf
\If{$is\_valid$}
\State evaluate\_and\_evolve\_dse($unroll\_factors$, $dse$)
\Else
\State evolve\_dse($unroll\_factors$, $dse$)
\EndIf
\Until{is\_converged($dse$) \textbf{or} is\_early\_terminated($dse$)}
\State apply\_unrolling($n$, get\_final\_unroll\_factors($dse$))
\end{algorithmic}
\end{algorithm}

\textbf{Step~(4) Node Parallelization.}
Dataflow nodes are parallelized in the order determined by step~(2). Algorithm~\ref{alg:node_parallelization} shows the pseudo-code of node parallelization. For each node \emph{n}, we first query whether any nodes connected with it have been parallelized (line 2). If so, each connected node's unroll factors are multiplied with the scaling map and then permutated with the permutation map generated in step~(1) (lines 3 to 8). The processed unroll factors are recorded in a \emph{constraints\_list} to constrain the intra-node DSE. Meanwhile, in line 9, we also get the parallel factor of node \emph{n} generated in step~(3) to constrain the overall parallelism of \emph{n}.

With all the constraints generated, lines 10 to 23 of Algorithm~\ref{alg:node_parallelization} illustrate a simplified intra-node DSE for node \emph{n}. In each iteration of exploration, the DSE engine will propose new unroll factors for evaluation. However, the proposed factors could fail to meet the constraints in two cases: (1)~If any of the unroll factors are mutually indivisible with the corresponding constraint (lines 13 to 16); or (2)~the overall parallelism exceeds the pre-calculated parallel factor (lines 17 to 18). Case~(1) can cause unaligned inter-node memory access behavior, while case~(2) can cause imbalanced dataflow execution, both leading to sub-optimal dataflow efficiency. If all the constraints are fulfilled, HIDA employs a quality of results (QoR) estimator from~\cite{hpca2022scalehls} to evaluate the performance and resource utilization of the proposed factors (line 20). Then, the evaluation results or failure message are passed to the DSE and evolve the exploration. Finally, in line 23, we terminate the DSE if the results have converged or met the early termination criteria. The proposed algorithm can identify the Pareto frontier in the local design space and selects the best design point under the imposed constraints.


\begin{table}[t]
    \small
    \centering
    \caption{Array partition results of Listing~\ref{code:dataflow_cpp}.}
    \label{tab:array_factors}
    \vspace{-4pt}
    \setlength\tabcolsep{3pt}
    \begin{tabular}{ccccccccccc}
        \toprule
        \multirow{2}{*}[-2pt]{\textbf{Array}} &
        \multicolumn{4}{c}{\textbf{Array Partition Factors}} &
        \multicolumn{4}{c}{\textbf{Bank Number}} \\
        \cmidrule(lr){2-5} \cmidrule(lr){6-9}
        & \textbf{IA+CA} & \textbf{IA} & \textbf{CA} & \textbf{Naive} & \textbf{IA+CA} & \textbf{IA} & \textbf{CA} & \textbf{Naive} \\
        \midrule
        \textbf{A} & [8, 1] & [8, 2] & [8, 4] & [8, 8] & 8 & 16 & 32 & 64 \\
        \textbf{B} & [1, 8] & [2, 8] & [4, 8] & [8, 8] & 8 & 16 & 32 & 64 \\
        \textbf{C} & [4, 8] & [4, 8] & [4, 8] & [4, 8] & 32 & 32 & 32 & 32 \\
        \bottomrule
    \end{tabular}
\end{table}

\subsubsection{Discussion}
We summarize the results of IA+CA, IA-only, CA-only, and naive parallelization in Table~\ref{tab:node_factors}, where the naive solution applies the maximum parallel factor 32 to all dataflow nodes. One can observe our IA+CA approach achieves the best unroll factors, eventually leading to the least computation resource utilization. Meanwhile, IA and CA can also reduce memory resource utilization. Table~\ref{tab:array_factors} shows the array partition results. Array partitioning is an HLS technique that divides a large array into smaller sub-arrays to enable parallel access. One can observe that our IA+CA approach achieves the lowest number of banks for arrays appearing in Listing~\ref{code:dataflow_cpp}, resulting in the lowest memory utilization. For this small example, the margin can already reach 8$\times$ on arrays \emph{A} and \emph{B} compared to the naive solution. For large-scale dataflow applications, the parallelization solution can determine whether the overall solution is scalable; an ablation study is conducted in Section~\ref{sec:ablation}.

\section{Evaluation}
\label{sec:evaluation}

To evaluate HIDA, we use FPGAs as the target platform and perform two sets of experiments using C++ and PyTorch inputs and an ablation study on a ResNet-18 model. As depicted in Figure~\ref{fig:framework}, AMD Vitis HLS 2022.1~\cite{vitishls2022userguide} is used for generating RTL code. All reported performances and resource utilization are collected from the synthesis results of Vitis HLS.

\begin{table*}
    \small
    \centering
    \caption{Evaluation results for C++ kernels. \emph{ScaleHLS} designs are automatically generated by~\cite{hpca2022scalehls}. \emph{SOFF} results are ported from their paper~\cite{jo2020soff}, which compared with SDAccel~(previous name of Vitis). \emph{Vitis} designs are solely optimized by Vitis HLS.}
    \label{tab:cpp_results}
    \vspace{-4pt}
    \setlength\tabcolsep{5pt}
    \begin{tabular}{ccccccccc}
        \toprule
        \multirow{2}{*}[-2pt]{\textbf{Kernel}} &
        \multirow{2}{*}[3pt]{\tabincell{c}{\textbf{HIDA} \\ \textbf{Compile} \\ \textbf{Time (s)}}} &
        \multirow{2}{*}[-2pt]{\tabincell{c}{\textbf{LUT} \\ \textbf{Number}}} &
        \multirow{2}{*}[-2pt]{\tabincell{c}{\textbf{FF} \\ \textbf{Number}}} &
        \multirow{2}{*}[-2pt]{\tabincell{c}{\textbf{DSP} \\ \textbf{Number}}} &
        \multicolumn{4}{c}{\textbf{Throughput (Samples/s)*}} \\
        \cmidrule(lr){6-9}
        & & & & &
        \textbf{HIDA} &
        \tabincell{c}{\textbf{ScaleHLS~\cite{hpca2022scalehls}}} & \tabincell{c}{\textbf{SOFF~\cite{jo2020soff}}} &
        \tabincell{c}{\textbf{Vitis~\cite{vitishls2022userguide}}} \\
        \midrule
        \textbf{2mm} & 0.65 & 38.8k & 27.4k & 269 & 239.22 & 122.39 (1.95$\times$) & 30.67 (7.80$\times$) & 1.23 (194.88$\times$) \\
        \textbf{3mm} & 0.79 & 38.7k & 27.8k & 243 & 175.43 & 92.33 (1.90$\times$) & - & 1.04 (167.99$\times$) \\
        \textbf{atax} & 2.06 & 44.6k & 34.6k & 260 & 1,021.39 & 932.26 (1.10$\times$) & 2,173.17 (0.47$\times$) & 103.18 (9.90$\times$) \\
        \textbf{bicg} & 0.72 & 16.0k & 15.1k & 61 & 2,869.69 & 2,869.61 (1.00$\times$) & 2,295.75 (1.25$\times$) & 104.19 (27.54$\times$) \\
        \textbf{correlation} & 0.91 & 14.5k & 12.3k & 66 & 67.33 & 59.77 (1.13$\times$) & 3.96 (16.99$\times$) & 1.32 (50.97$\times$) \\
        \textbf{gesummv} & 0.60 & 34.2k & 22.8k & 232 & 31,685.68 & 31,685.68 (1.00$\times$) & 3,466.70 (9.14$\times$) & 266.65 (118.83$\times$) \\
        \textbf{jacobi-2d} & 1.98 & 91.4k & 56.6k & 352 & 257.27 & 128.63 (2.00$\times$) & - & 2.71 (94.95$\times$) \\
        \textbf{mvt} & 0.42 & 23.8k & 16.5k & 162 & 9,979.04 & 4,989.02 (2.00$\times$) & 870.01 (11.47$\times$) & 62.13 (160.62$\times$) \\
        \textbf{seidel-2d} & 3.59 & 5.5k & 2.5k & 4 & 0.14 & 0.14 (1.00$\times$) & - & 0.11 (1.28$\times$) \\
        \textbf{symm} & 1.05 & 14.9k & 9.5k & 74 & 2.62 & 2.62 (1.00$\times$) & - & 2.02 (1.29$\times$) \\
        \textbf{syr2k} & 0.69 & 14.3k & 12.8k & 78 & 27.68 & 27.67 (1.00$\times$) & - & 1.44 (19.23$\times$) \\
        \midrule
        \textbf{Geo. Mean} & \textbf{0.99} & & & & & \textbf{1.29$\times$} & \textbf{4.49$\times$} & \textbf{31.08$\times$} \\
        \bottomrule \vspace{-8pt} \\
        \multicolumn{9}{l}{* Numbers in () show throughput improvements of HIDA over others.}
    \end{tabular}
\end{table*}

\subsection{C++ Kernels Evaluation}

\textbf{Experiment Settings.}
We evaluate HIDA with a set of C++ benchmarks from PolyBench~\cite{pouchet2012polybench}. The benchmarks cover multiple categories, including blas routines~(gesummv, symm, and syr2k), linear algebra kernels~(2mm, 3mm, atax, bicg, and mvt), data mining~(correlation), and stencils~(jacobi-2d and seidel-2d). The target platform is AMD-Xilinx ZU3EG FPGA. Table~\ref{tab:cpp_results} shows the evaluation results. Although Vitis HLS can automatically apply optimizations such as loop \texttt{pipeline}, it cannot conduct complex dataflow analysis and optimizations. As a result, HIDA achieves 31.08$\times$ higher throughput on average over Vitis HLS.

\textbf{Comparison with Previous Works.} 
Compared with the state-of-the-art~(SOTA) HLS optimization framework ScaleHLS~\cite{hpca2022scalehls}, and another HLS framework SOFF~\cite{jo2020soff}, HIDA achieved 1.29$\times$ and 4.49$\times$ higher throughput, respectively. We observed that for single-loop kernels~(bicg, gesummv, seidel-2d, symm, and syr2k), the performance of HIDA was on par with ScaleHLS due to these kernels not presenting any dataflow optimization opportunities. For the multi-loop kernels, HIDA outperforms ScaleHLS due to dataflow optimizations. When only considering multi-loop kernels, HIDA achieves 1.57$\times$ higher throughput than ScaleHLS. We concluded that the dataflow scheduling and parallelization problems are pervasive based on the evaluation results. Thus, HIDA-OPT can better optimize these kernels, ultimately leading to an increased performance.

\begin{table*}
    \small
    \centering
    \caption{Evaluation results for PyTorch models. \emph{DNNBuilder} results are directly from their paper~\cite{zhang2018dnnbuilder}. To make fair comparison, we constrained the FPGA resources to the same with DNNBuilder. \emph{ScaleHLS} designs are automatically generated by~\cite{hpca2022scalehls}.}
    \label{tab:dnn_results}
    \vspace{-4pt}
    \setlength\tabcolsep{5pt}
    \begin{tabular}{ccccccccccccc}
        \toprule
        \multirow{3}{*}[-2pt]{\textbf{Model}} &
        \multirow{3}{*}[-2pt]{\tabincell{c}{\textbf{HIDA} \\ \textbf{Compile} \\ \textbf{Time (s)}}} &
        \multirow{3}{*}[-2pt]{\tabincell{c}{\textbf{LUT} \\ \textbf{Number}}} &
        \multirow{3}{*}[-2pt]{\tabincell{c}{\textbf{DSP} \\ \textbf{Number}}} &
        \multicolumn{3}{c}{\tabincell{c}{\textbf{Throughput (Samples/s)*}}} &
        \multicolumn{3}{c}{\tabincell{c}{\textbf{DSP Efficiency*}}} \\
        \cmidrule(lr){5-7} \cmidrule(lr){8-10}
        & & & &
        \textbf{HIDA} &
        \tabincell{c}{\textbf{DNNBuilder} \\ \textbf{\cite{zhang2018dnnbuilder}}} &
        \tabincell{c}{\textbf{ScaleHLS} \\ \textbf{\cite{hpca2022scalehls}}} &
        \textbf{HIDA} &
        \tabincell{c}{\textbf{DNNBuilder} \\ \textbf{\cite{zhang2018dnnbuilder}}} &
        \tabincell{c}{\textbf{ScaleHLS} \\ \textbf{\cite{hpca2022scalehls}}} \\
        \midrule
        \textbf{ResNet-18} & 83.1 & 142.1k & 667 & 45.4 & - & 3.3 (13.88$\times$) & 73.8\% & - & 5.2\% (14.24$\times$) \\
        \textbf{MobileNet} & 110.8 & 132.9k & 518 & 137.4 & - & 15.4 (8.90$\times$) & 75.5\% & - & 9.6\% (7.88$\times$) \\
        \textbf{ZFNet} & 116.2 & 103.8k & 639 & 90.4 & 112.2 (0.81$\times$) & - & 82.8\% & 79.7\% (1.04$\times$) & - \\
        \textbf{VGG-16} & 199.9 & 266.2k & 1118 & 48.3 & 27.7 (1.74$\times$) & 6.9 (6.99$\times$) & 102.1\% & 96.2\% (1.06$\times$) & 18.6\% (5.49$\times$) \\
        \textbf{YOLO} & 188.2 & 202.8k & 904 & 33.7 & 22.1 (1.52$\times$) & - & 94.3\% & 86.0\% (1.10$\times$) & - \\
        \textbf{MLP} & 40.9 & 21.0k & 164 & 938.9 & - & 152.6 (6.15$\times$) & 90.0\% & - & 17.6\% (5.10$\times$) \\
        \midrule
        \textbf{Geo. Mean} & \textbf{108.7} & & & & \textbf{1.29$\times$} & \textbf{8.54$\times$} & & \textbf{1.07$\times$} & \textbf{7.49$\times$} \\
        \bottomrule \vspace{-8pt} \\
        \multicolumn{10}{l}{* Numbers in () show throughput/DSP efficiency improvements of HIDA over others.}
    \end{tabular}
\end{table*}

\subsection{PyTorch Models Evaluation}

\textbf{Experiment Settings.}
We evaluate HIDA with a set of deep neural network~(DNN) benchmarks written in PyTorch to understand its performance on large-scale dataflow applications. The benchmarks cover multiple categories of DNN models, including image classification~(ResNet-18~\cite{he2016deep}, MobileNet~\cite{howard2017mobilenets}, ZFNet~\cite{zeiler2014visualizing}, and VGG-16~\cite{simonyan2014very}), object detection~(YOLO~\cite{redmon2016you}), and fully-connected networks~(MLP). The optimization for these models exhibit significant variations under dataflow setting, owing to the distinct layer types and interconnections. The targeted platform is one super logic region~(SLR) of an AMD-Xilinx VU9P FPGA. Table~\ref{tab:dnn_results} shows the evaluation results. Even for these complicated DNN models, HIDA only takes 108.7 seconds on average to compile them into dataflow implementations.

\begin{figure}[t]
    \centering
    \includegraphics[width=\linewidth]{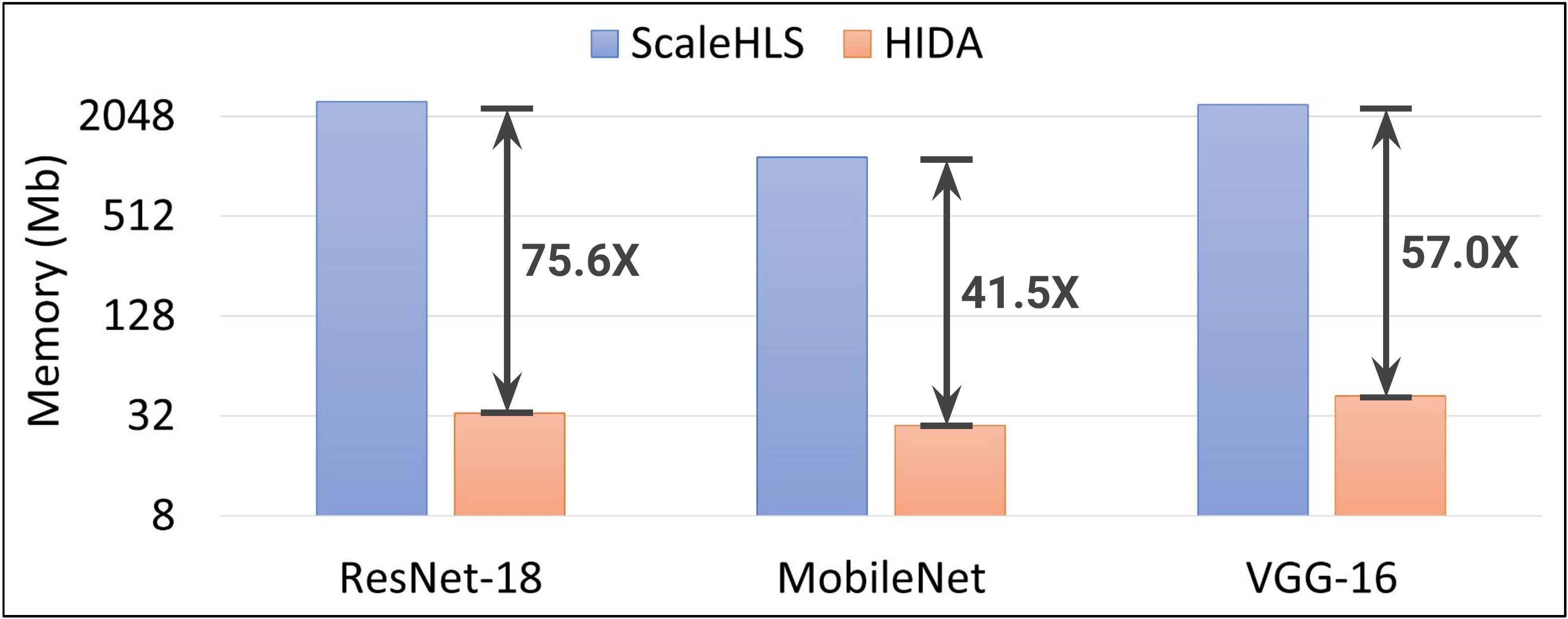}
    \vspace{-16pt}
    \caption{Memory utilization compared with ScaleHLS~\cite{hpca2022scalehls}.}
    \label{fig:memory_compare}
\end{figure}

\textbf{Comparison with Previous Works.} 
Again, we compare HIDA with ScaleHLS~\cite{hpca2022scalehls}, where we observe an 8.54$\times$ higher throughput. The throughput gains are much more significant than the C++ kernels due to large DNN models exposing more opportunities for HIDA to optimize the dataflow architecture. For ZFNet and YOLO, ScaleHLS cannot produce results due to the DNNs having irregular convolution sizes and high-resolution inputs, respectively, demonstrating the superior flexibility and scalability of HIDA. For the four benchmarks supported by ScaleHLS, we use DSP efficiency to compare the two frameworks, calculated as:
\begin{equation}
    Efficiency_{DSP} = \frac{Throughput \times OPs}{Number_{DSP} \times Frequency},
\end{equation}
where $OPs$ denotes the number of multiply-accumulate~(MAC) operations per sample of the DNN, \emph{Throughput} is samples per second, and $Frequency$ denotes the clock frequency constant at 200MHz for both ScaleHLS and HIDA. DSP efficiency is a common metric for comparing the efficiency of DNN accelerators across different platforms or frameworks. A 100\% of DSP efficiency indicates all instantiated DSPs in the accelerator continuously operating without stalling. HIDA achieves 7.49$\times$ higher DSP efficiency than ScaleHLS on average and 14.24$\times$ for ResNet-18. We attribute the much higher efficiency for ResNet-18 to HIDA's ability to optimize shortcut data paths. For VGG-16, we observe an over 100\% DSP efficiency, attributing the excess percentage to the back-end RTL generator where MAC operations can be instantiated with LUTs when resources are abundant.

Apart from the throughput and efficiency improvements, we also observe substantial on-chip memory reduction by HIDA compared to ScaleHLS. As Figure~\ref{fig:memory_compare} shows, HIDA can reduce memory utilization by 41.5$\times$ to 75.6$\times$ due to several factors: (1)~HIDA can leverage loop tiling and local buffer creation to only cache small tiles of intermediate results while enabling the dataflow execution. In comparison, ScaleHLS must keep all intermediate results on-chip due to the lack of external memory access support. (2)~The IA+CA parallelization can drastically reduce the buffer sizes. In summary, HIDA can utilize computation and memory resources more efficiently and achieve substantial throughput improvements on DNN models compared to SOTA frameworks.

\textbf{Comparison with Dedicated DNN Accelerator.} In addition to the comparison with HLS optimization frameworks, we further compare HIDA with a dedicated DNN acceleration framework, DNNBuilder~\cite{zhang2018dnnbuilder}. DNNBuilder has RTL-based and human-designed DNN IPs and can enable the dataflow execution of all the instantiated IPs to achieve SOTA throughput and efficiency on FPGAs. As shown in Table~\ref{tab:dnn_results}, HIDA achieves 1.29$\times$ and 1.07$\times$ higher throughput and DSP efficiency compared to DNNBuilder, which already has an extremely high DSP efficiency. Note that DNNBuilder doesn't support ResNet-18 and MobileNet due to its lack of support for shortcut paths and depthwise convolutions. Through this comparison, we demonstrate the productivity and performance of HIDA outperforming a dedicated DNN acceleration framework. Additionally, we demonstrate the flexibility of HIDA, which can automatically adapt to a wide range of computational patterns.

\subsection{Ablation Study on ResNet-18}
\label{sec:ablation}

\begin{figure}[t]
    \centering
    \includegraphics[width=\linewidth]{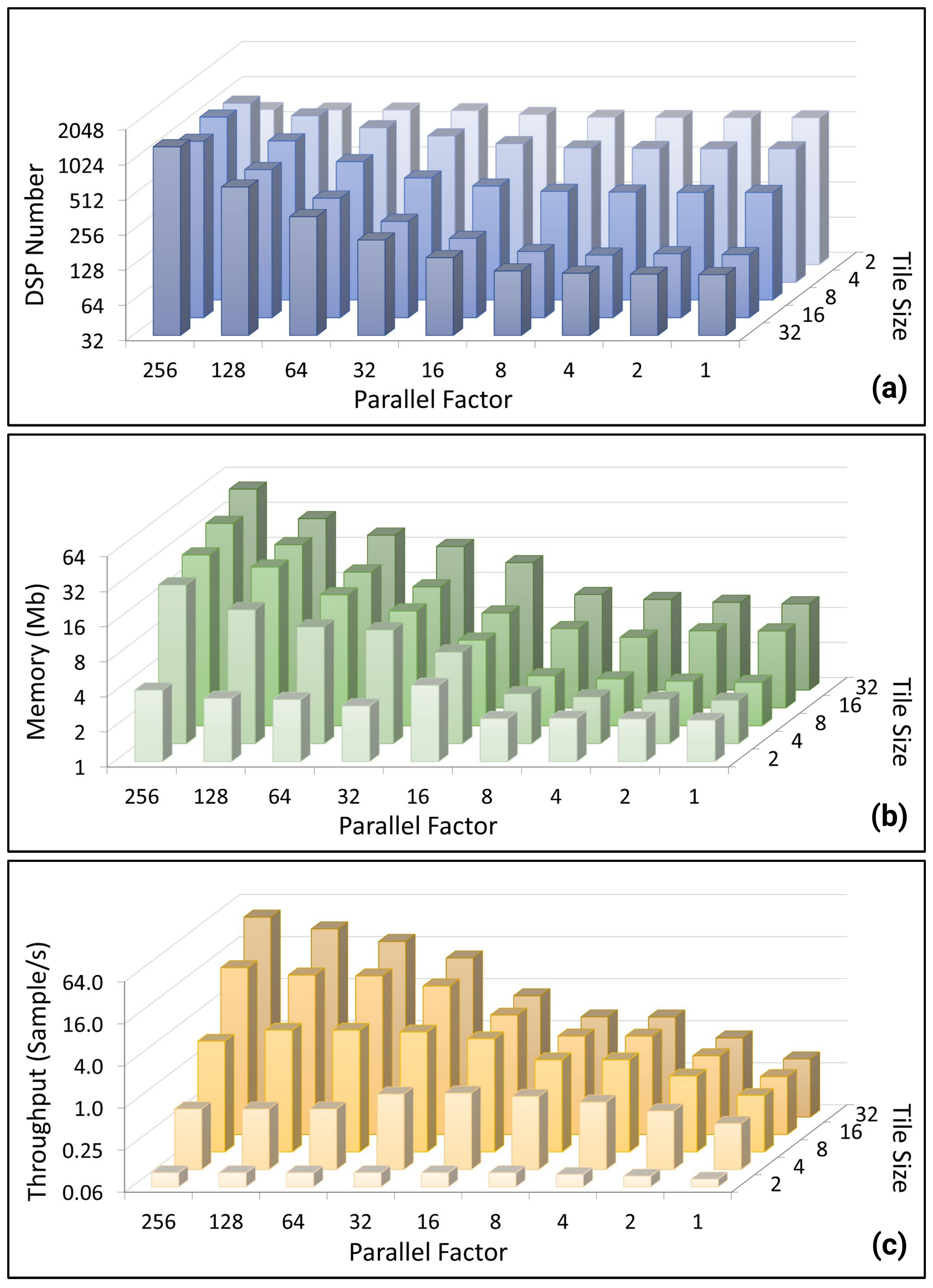}
    \vspace{-16pt}
    \caption{Parallel factor and tile size ablation on ResNet-18.}
    \label{fig:parallel_tile_ablation}
\end{figure}

\textbf{Parallel Factor and Tile Size Ablation.}
To understand the scalability of HIDA, we perform an ablation study on ResNet-18 by sweeping the maximum parallel factor from 1 to 256 and tile size from 2 to 32. Three metrics~(DSP utilization, memory utilization, and throughput) are measured for each combination of parallel factor and tile size. Figure~\ref{fig:parallel_tile_ablation} shows the ablation study results. Overall, as the parallel factor increases, all three metrics increase as expected, showing promising \emph{scalability} of HIDA. Meanwhile, the overall trend predictably showed more memory instances being utilized as we increased the tile size. However, we also observed some interesting findings summarized as follows:
\begin{itemize}[leftmargin=*]
    \item \emph{Small tile can drastically increase DSP utilization.} Counter-intuitively, the number of DSPs increases when the tile size decreases, highlighted by the data point with a parallel factor of 1 and a tile size of 2, instantiating 518 DSPs. Analysis shows that due to small tiles needing fine-grained control of memory accesses, a large number of DSPs are used for address calculations instead of actual computations.
    \item \emph{Smaller tile may not reduce memory utilization.} For small parallel factors~(1 to 8) and small tile sizes~(2 to 8), the memory utilization does not change noticeably within the range. Analysis shows that as the minimum memory instance is BRAM~(block RAM) for FPGAs, increasing parallel factor or tile size may not demand more BRAM instances as long as the current BRAMs are large enough to hold the data tiles and can provide enough bandwidth.
    \item \emph{Throughput and tile size may positively correlate.} Ideally, different tile sizes should not affect the throughput. However, experimental results disagree: throughput increases with tile size being larger, especially for large parallel factors~(32 to 256). Analysis shows two main reasons for the observed behavior: (1)~Small tile sizes cannot provide sufficient bandwidth, resulting in a degradation of the level of parallelism.(2)~Small tile sizes cannot provide sufficient burst length for consecutive external memory access, leading to poor external memory efficiency.
\end{itemize}

\begin{figure}[t]
    \centering
    \includegraphics[width=\linewidth]{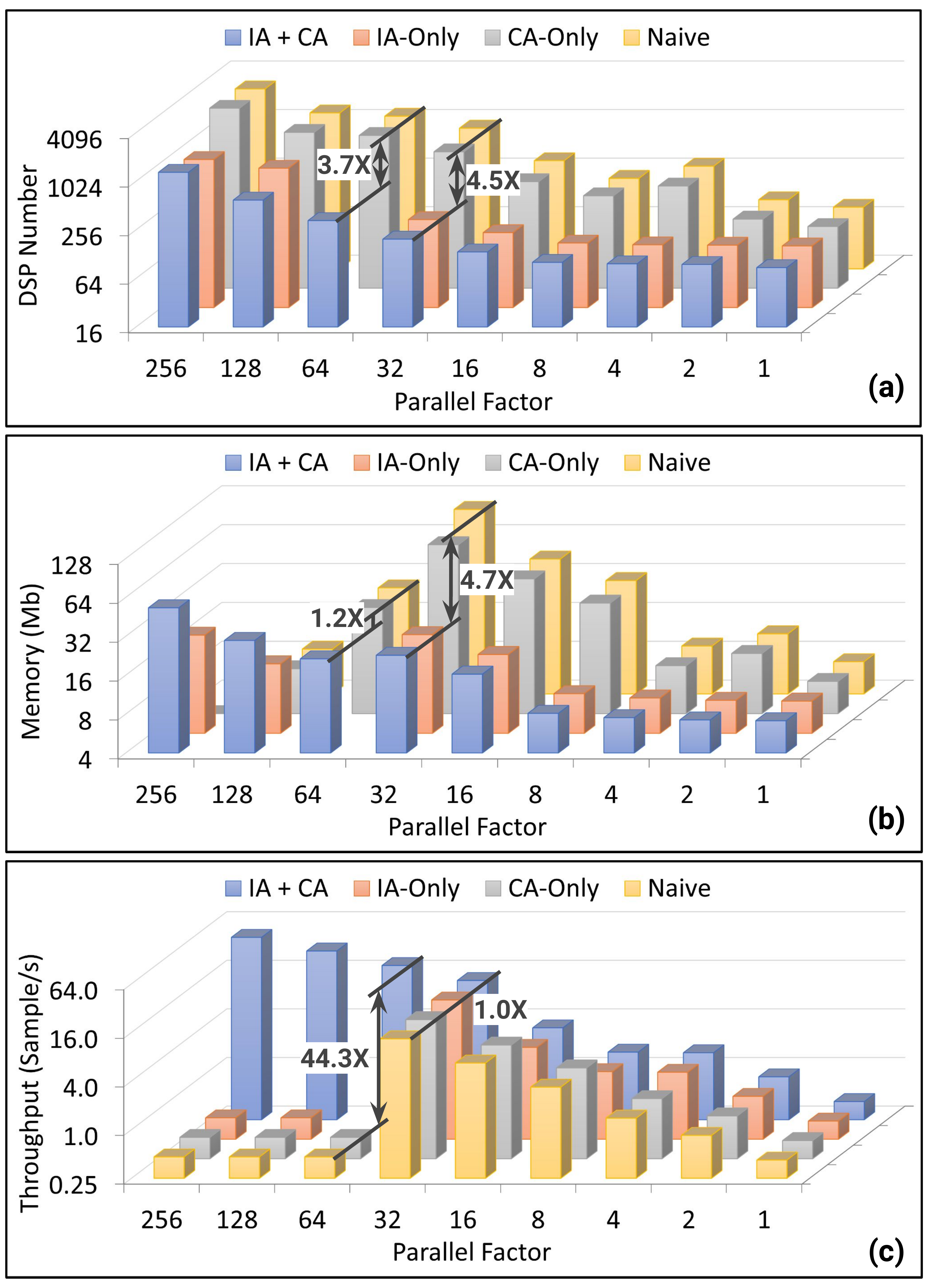}
    \vspace{-16pt}
    \caption{Intensity-aware~(IA) and connection-aware~(CA) dataflow parallelization ablation on ResNet-18.}
    \label{fig:optimization_ablation}
\end{figure}

\textbf{Node Parallelization Ablation.}
Section~\ref{sec:opt} introduces IA and CA approaches to drive the dataflow parallelization. To better understand the performance of the two methods on large-scale applications, we conduct an ablation study on ResNet-18. In this study, we set four groups of experiments~(IA+CA, IA-only, CA-only, and Naive) for comparison, and for each group, we sweep the maximum parallel factor from 1 to 256. As expected, the four groups show a similar trend in DSP numbers - they increase proportionally with the parallel factor. However, the IA+CA group shows a drastically different trend in memory and throughput compared with the other groups - only IA+CA scales well when we increase the parallel factor. For instance, for the data samples with a parallel factor of 64, compared with the other groups, IA+CA utilizes 3.7$\times$ less DSP and 1.2$\times$ less memory, yet achieves 44.3$\times$ better throughput. After studying the generated accelerators, we found for all the other groups except IC+CA, the compiler generates overly-complicated control logics due to the mismatch between node unroll factors and memory layouts, ultimately falling back to flawed designs. Meanwhile, even when all groups can be scaled well, IA+CA shows substantially better resource utilization efficiency than other groups. For instance, for the data samples with a parallel factor of 32, compared with the other groups, IA+CA performs on par in throughput but utilizes 4.5$\times$ less DSP and 4.7$\times$ less memory. Through these comparisons, we demonstrate the superior scalability and efficiency of our proposed IA+CA dataflow parallelization approach.

\section{Related Works}
\label{sec:related_works}

\subsection{MLIR Infrastructure}
MLIR~\cite{lattner2021mlir, mlir2023github} is a compiler infrastructure with multiple levels of representation, allowing users to tailor domain-specific compilers. MLIR also provides commonly used IRs and optimizations for different domains, including tensor~\cite{tensor2023document}, linear algebra~\cite{linalg2023document}, and loop~\cite{affine2023document}. As MLIR is an infrastructure for compilers, it does not natively support hardware-oriented compilation; however, many related compilers have been developed on top of MLIR. IREE~\cite{iree2023github} provides compilation and runtime support for hardware accelerators like GPU. CIRCT~\cite{circt2023github} is a toolchain for circuit design and optimization, supporting HLS, HDL-generation languages~\cite{bachrach2012chisel}, and HDL.

\subsection{HLS Languages}
Different domain-specific languages~(DSL) have been proposed to improve the productivity and performance of commercial HLS tools. TAPA~\cite{chi2021extending, guo2022tapa} introduced specialized interfaces to enable dataflow and efficient external memory access, but didn't provide solution for dataflow DSE. As a result, TAPA requires users to make many design decisions, such as task parallelization and buffer implementation. Spatial~\cite{koeplinger2018spatial} abstracted interfaces for describing the control, memory, and I/O structures, enabling auto-tuning-based DSE. Dahlia~\cite{nigam2020predictable} proposed an affine-based approach to improve the predictability of HLS designs. Aetherling~\cite{durst2020type} proposed a strong type system to tackle the fine-grained scheduling and DSE of hardware design. Another cluster of works extends existing language syntax, such as Python, to further raise the abstraction level of HLS designs, including DaCe~\cite{ben2019stateful} and PyLog~\cite{huang2021pylog}. These languages are orthogonal to HIDA, which attempts to address the dataflow optimization problem with compiler techniques. As future works, they can be integrated as front-ends to further improve the productivity of HIDA.

\textbf{HeteroCL.}
HeteroCL~\cite{lai2019heterocl} and HeteroFlow~\cite{xiang2022heteroflow} decoupled the algorithmic description and optimization of HLS designs by providing a versatile set of computation, data type, and data movement customization primitives. HeteroCL incorporated third-party frameworks for automated HLS optimization, including PolySA~\cite{cong2018polysa} for systolic arrays and SODA~\cite{chi2018soda} for stencil applications. However, these automated optimizations are intra-task and domain-specific. For inter-task optimizations and computation patterns that are not covered by third-party frameworks, designers must make every design decision empirically. Meanwhile, HeteroCL cannot directly take PyTorch as an input - designers must rewrite the PyTorch model in HeteroCL DSL.

\subsection{HLS Compilers}
Existing works have explored different approaches for the fine-grained scheduling problem in HLS, including static scheduling~(Spatial~\cite{koeplinger2018spatial}, SOFF~\cite{jo2020soff}, and Calyx~\cite{nigam2021compiler}), dynamic scheduling~(Dynamatic~\cite{josipovic2018dynamically} and TAPAS~\cite{margerm2018tapas}), and static-dynamic hybrid scheduling~(DASS~\cite{cheng2020combining} and Hector~\cite{xu2022hector}). Fine-grained scheduling deals with operator-level parallelism, such as multipliers, instead of task-level and thus is very different from the problems we addressed in HIDA. For instance, fine-grained scheduling does not consider the intra-task parallelization. We have seen a large amount of HLS optimization tools, including LLVM-based~(Merlin~\cite{cong2016source}, COMBA~\cite{zhao2017comba}, and HPVM2FPGA~\cite{ejjeh2022hpvm2fpga}) and MLIR-based~(ScaleHLS~\cite{hpca2022scalehls}, POLSCA~\cite{zhao2022polsca}, and SODA-Opt~\cite{agostini2022mlir}). However, as discussed in Section~\ref{sec:introduction}, these tools either did not consider dataflow during the compilation or are limited in dataflow-oriented optimizations. Another alternative solution is leveraging existing multi-core CPU or CGRA compilers, such as PolyMage~\cite{mullapudi2015polymage}, Tapir~\cite{schardl2017tapir}, Unified Buffer~\cite{liu2023unified}, and Revet~\cite{rucker2023revet}, for the purpose of HLS. However, the dataflow optimizations are not compatible with these compilers due to the fundamentally different programming model and memory hierarchy of HLS-based dataflow accelerators.

\textbf{CIRCT.} Handshake dialect~\cite{handshake2023document} is a CIRCT dialect implementing the elastic circuit components and dynamic scheduling algorithms proposed in Dynamatic~\cite{josipovic2018dynamically}. Coarse-grained tasks are abstracted as \emph{function}s in the handshake dialect, where scalar intermediate results are passed between different functions through handshaking FIFOs. However, for tensor intermediate results, the handshake dialect adopts a load-store queue-based~\cite{josipovic2017out} shared memory model instead of the dataflow model for inter-function communication, drawing a line between the two. Elastic silicon interconnect~(ESI) dialect~\cite{esi2023document} is another CIRCT dialect aiming to provide type-safe and latency-insensitive interface abstraction for FPGA/ASIC design. In contrast, HIDA aims to tackle the DSE problem of fine-grained and coarse-grained scheduling, which is orthogonal to the mission of the ESI dialect.

\subsection{DNN Compilers}
DNN layer fusion algorithms have been studied in recent years~\cite{alwani2016fused, zhao2018deepthings} to reduce the layer-wise communication cost of DNN training or inference. In contrast, HIDA’s task fusion is a general-purpose algorithm for dataflow applications in different domains. The patterns proposed by~\cite{alwani2016fused, zhao2018deepthings} can be implemented and plugged into HIDA. The scheduling problem of cache-based or scratchpad-based DNN accelerators has also been thoroughly studied, such as in Eyeriss~\cite{chen2016eyeriss}, Diannao~\cite{chen2014diannao}, and Timeloop~\cite{parashar2019timeloop}, either from an architecture or compiler perspective. However, the scheduling of dataflow-based DNN accelerators is still under intensive study~\cite{zhuang2023charm, rucker2023revet, zhao2023sigma}. The intention to balance the dataflow pipeline while enabling buffer sharing or streaming between layer instances ~(either a single DNN layer, a fused layer, or a decomposed layer) significantly complicates the scheduling problem. The reason is, as we mentioned in Section~\ref{sec:parallelization}, the local optimality of each layer cannot lead to the global optimal dataflow architecture. Although existing works, such as FINN~\cite{umuroglu2017finn} and DNNBuilder~\cite{zhang2018dnnbuilder}, explored the problem to some extent, their framework can only target a subset of DNNs; for instance, DNNBuilder only supports CNNs.

\section{Conclusion}
\label{sec:conclusion}
In this paper, we propose HIDA, an HLS framework that can systematically transform an algorithmic description into an efficient dataflow implementation. We propose a two-level dataflow representation, HIDA-IR, and a hierarchical dataflow optimizer, HIDA-OPT, significantly improving the productivity, performance, and scalability of HLS dataflow accelerators. To demonstrate the performance of HIDA, we evaluate a set of DNN models and C++ kernels, where HIDA outperforms the existing SOTA hand-tuned RTL-based DNN accelerator and compilation-based HLS frameworks. We hope that the HIDA framework will serve as a new open infrastructure for future dataflow architectural research, allowing researchers to explore the vast design space effectively.

\begin{acks}
We thank all anonymous reviewers and Adrian Sampson of Cornell University for their valuable feedback and suggestions. This work is supported in part by AMD Center of Excellence at UIUC, AMD Heterogeneous Adaptive Compute Cluster (HACC) initiative, NSF 2117997 grant through the A3D3 institute, and Semiconductor Research Corporation (SRC) 2023-CT-3175 grant.
\end{acks}

\bibliographystyle{ACM-Reference-Format}
\bibliography{references}


\begin{thebibliography}{87}


\ifx \showCODEN    \undefined \def \showCODEN     #1{\unskip}     \fi
\ifx \showDOI      \undefined \def \showDOI       #1{#1}\fi
\ifx \showISBNx    \undefined \def \showISBNx     #1{\unskip}     \fi
\ifx \showISBNxiii \undefined \def \showISBNxiii  #1{\unskip}     \fi
\ifx \showISSN     \undefined \def \showISSN      #1{\unskip}     \fi
\ifx \showLCCN     \undefined \def \showLCCN      #1{\unskip}     \fi
\ifx \shownote     \undefined \def \shownote      #1{#1}          \fi
\ifx \showarticletitle \undefined \def \showarticletitle #1{#1}   \fi
\ifx \showURL      \undefined \def \showURL       {\relax}        \fi
\providecommand\bibfield[2]{#2}
\providecommand\bibinfo[2]{#2}
\providecommand\natexlab[1]{#1}
\providecommand\showeprint[2][]{arXiv:#2}

\bibitem[Agostini et~al\mbox{.}(2022)]%
        {agostini2022mlir}
\bibfield{author}{\bibinfo{person}{Nicolas~Bohm Agostini}, \bibinfo{person}{Serena Curzel}, \bibinfo{person}{Vinay Amatya}, \bibinfo{person}{Cheng Tan}, \bibinfo{person}{Marco Minutoli}, \bibinfo{person}{Vito~Giovanni Castellana}, \bibinfo{person}{Joseph Manzano}, \bibinfo{person}{David Kaeli}, {and} \bibinfo{person}{Antonino Tumeo}.} \bibinfo{year}{2022}\natexlab{}.
\newblock \showarticletitle{An MLIR-based Compiler Flow for System-Level Design and Hardware Acceleration}. In \bibinfo{booktitle}{\emph{Proceedings of the 41st IEEE/ACM International Conference on Computer-Aided Design}}. \bibinfo{pages}{1--9}.
\newblock


\bibitem[Alwani et~al\mbox{.}(2016)]%
        {alwani2016fused}
\bibfield{author}{\bibinfo{person}{Manoj Alwani}, \bibinfo{person}{Han Chen}, \bibinfo{person}{Michael Ferdman}, {and} \bibinfo{person}{Peter Milder}.} \bibinfo{year}{2016}\natexlab{}.
\newblock \showarticletitle{Fused-layer CNN accelerators}. In \bibinfo{booktitle}{\emph{2016 49th Annual IEEE/ACM International Symposium on Microarchitecture (MICRO)}}. IEEE, \bibinfo{pages}{1--12}.
\newblock


\bibitem[Bachrach et~al\mbox{.}(2012)]%
        {bachrach2012chisel}
\bibfield{author}{\bibinfo{person}{Jonathan Bachrach}, \bibinfo{person}{Huy Vo}, \bibinfo{person}{Brian Richards}, \bibinfo{person}{Yunsup Lee}, \bibinfo{person}{Andrew Waterman}, \bibinfo{person}{Rimas Avi{\v{z}}ienis}, \bibinfo{person}{John Wawrzynek}, {and} \bibinfo{person}{Krste Asanovi{\'c}}.} \bibinfo{year}{2012}\natexlab{}.
\newblock \showarticletitle{Chisel: constructing hardware in a scala embedded language}. In \bibinfo{booktitle}{\emph{Proceedings of the 49th Annual Design Automation Conference}}. \bibinfo{pages}{1216--1225}.
\newblock


\bibitem[Ben-Nun et~al\mbox{.}(2019)]%
        {ben2019stateful}
\bibfield{author}{\bibinfo{person}{Tal Ben-Nun}, \bibinfo{person}{Johannes de Fine~Licht}, \bibinfo{person}{Alexandros~N Ziogas}, \bibinfo{person}{Timo Schneider}, {and} \bibinfo{person}{Torsten Hoefler}.} \bibinfo{year}{2019}\natexlab{}.
\newblock \showarticletitle{Stateful dataflow multigraphs: A data-centric model for performance portability on heterogeneous architectures}. In \bibinfo{booktitle}{\emph{Proceedings of the International Conference for High Performance Computing, Networking, Storage and Analysis}}. \bibinfo{pages}{1--14}.
\newblock


\bibitem[Benabderrahmane et~al\mbox{.}(2010)]%
        {benabderrahmane2010polyhedral}
\bibfield{author}{\bibinfo{person}{Mohamed-Walid Benabderrahmane}, \bibinfo{person}{Louis-No{\"e}l Pouchet}, \bibinfo{person}{Albert Cohen}, {and} \bibinfo{person}{C{\'e}dric Bastoul}.} \bibinfo{year}{2010}\natexlab{}.
\newblock \showarticletitle{The polyhedral model is more widely applicable than you think}. In \bibinfo{booktitle}{\emph{Compiler Construction: 19th International Conference, CC 2010, Held as Part of the Joint European Conferences on Theory and Practice of Software, ETAPS 2010, Paphos, Cyprus, March 20-28, 2010. Proceedings 19}}. Springer, \bibinfo{pages}{283--303}.
\newblock


\bibitem[Chen et~al\mbox{.}(2014)]%
        {chen2014diannao}
\bibfield{author}{\bibinfo{person}{Tianshi Chen}, \bibinfo{person}{Zidong Du}, \bibinfo{person}{Ninghui Sun}, \bibinfo{person}{Jia Wang}, \bibinfo{person}{Chengyong Wu}, \bibinfo{person}{Yunji Chen}, {and} \bibinfo{person}{Olivier Temam}.} \bibinfo{year}{2014}\natexlab{}.
\newblock \showarticletitle{Diannao: A small-footprint high-throughput accelerator for ubiquitous machine-learning}.
\newblock \bibinfo{journal}{\emph{ACM SIGARCH Computer Architecture News}} \bibinfo{volume}{42}, \bibinfo{number}{1} (\bibinfo{year}{2014}), \bibinfo{pages}{269--284}.
\newblock


\bibitem[Chen et~al\mbox{.}(2016)]%
        {chen2016eyeriss}
\bibfield{author}{\bibinfo{person}{Yu-Hsin Chen}, \bibinfo{person}{Tushar Krishna}, \bibinfo{person}{Joel~S Emer}, {and} \bibinfo{person}{Vivienne Sze}.} \bibinfo{year}{2016}\natexlab{}.
\newblock \showarticletitle{Eyeriss: An energy-efficient reconfigurable accelerator for deep convolutional neural networks}.
\newblock \bibinfo{journal}{\emph{IEEE journal of solid-state circuits}} \bibinfo{volume}{52}, \bibinfo{number}{1} (\bibinfo{year}{2016}), \bibinfo{pages}{127--138}.
\newblock


\bibitem[Cheng et~al\mbox{.}(2020)]%
        {cheng2020combining}
\bibfield{author}{\bibinfo{person}{Jianyi Cheng}, \bibinfo{person}{Lana Josipovic}, \bibinfo{person}{George~A Constantinides}, \bibinfo{person}{Paolo Ienne}, {and} \bibinfo{person}{John Wickerson}.} \bibinfo{year}{2020}\natexlab{}.
\newblock \showarticletitle{Combining dynamic \& static scheduling in high-level synthesis}. In \bibinfo{booktitle}{\emph{Proceedings of the 2020 ACM/SIGDA International Symposium on Field-Programmable Gate Arrays}}. \bibinfo{pages}{288--298}.
\newblock


\bibitem[Chi et~al\mbox{.}(2018)]%
        {chi2018soda}
\bibfield{author}{\bibinfo{person}{Yuze Chi}, \bibinfo{person}{Jason Cong}, \bibinfo{person}{Peng Wei}, {and} \bibinfo{person}{Peipei Zhou}.} \bibinfo{year}{2018}\natexlab{}.
\newblock \showarticletitle{SODA: Stencil with optimized dataflow architecture}. In \bibinfo{booktitle}{\emph{2018 IEEE/ACM International Conference on Computer-Aided Design (ICCAD)}}. IEEE, \bibinfo{pages}{1--8}.
\newblock


\bibitem[Chi et~al\mbox{.}(2021)]%
        {chi2021extending}
\bibfield{author}{\bibinfo{person}{Yuze Chi}, \bibinfo{person}{Licheng Guo}, \bibinfo{person}{Jason Lau}, \bibinfo{person}{Young-kyu Choi}, \bibinfo{person}{Jie Wang}, {and} \bibinfo{person}{Jason Cong}.} \bibinfo{year}{2021}\natexlab{}.
\newblock \showarticletitle{Extending high-level synthesis for task-parallel programs}. In \bibinfo{booktitle}{\emph{2021 IEEE 29th Annual International Symposium on Field-Programmable Custom Computing Machines (FCCM)}}. IEEE, \bibinfo{pages}{204--213}.
\newblock


\bibitem[Cong et~al\mbox{.}(2016)]%
        {cong2016source}
\bibfield{author}{\bibinfo{person}{Jason Cong}, \bibinfo{person}{Muhuan Huang}, \bibinfo{person}{Peichen Pan}, \bibinfo{person}{Yuxin Wang}, {and} \bibinfo{person}{Peng Zhang}.} \bibinfo{year}{2016}\natexlab{}.
\newblock \showarticletitle{Source-to-source optimization for HLS}.
\newblock \bibinfo{journal}{\emph{FPGAs for Software Programmers}} (\bibinfo{year}{2016}), \bibinfo{pages}{137--163}.
\newblock


\bibitem[Cong et~al\mbox{.}(2011)]%
        {cong2011high}
\bibfield{author}{\bibinfo{person}{Jason Cong}, \bibinfo{person}{Bin Liu}, \bibinfo{person}{Stephen Neuendorffer}, \bibinfo{person}{Juanjo Noguera}, \bibinfo{person}{Kees Vissers}, {and} \bibinfo{person}{Zhiru Zhang}.} \bibinfo{year}{2011}\natexlab{}.
\newblock \showarticletitle{High-level synthesis for FPGAs: From prototyping to deployment}.
\newblock \bibinfo{journal}{\emph{IEEE Transactions on Computer-Aided Design of Integrated Circuits and Systems}} \bibinfo{volume}{30}, \bibinfo{number}{4} (\bibinfo{year}{2011}), \bibinfo{pages}{473--491}.
\newblock


\bibitem[Cong and Wang(2018)]%
        {cong2018polysa}
\bibfield{author}{\bibinfo{person}{Jason Cong} {and} \bibinfo{person}{Jie Wang}.} \bibinfo{year}{2018}\natexlab{}.
\newblock \showarticletitle{PolySA: Polyhedral-based systolic array auto-compilation}. In \bibinfo{booktitle}{\emph{2018 IEEE/ACM International Conference on Computer-Aided Design (ICCAD)}}. IEEE, \bibinfo{pages}{1--8}.
\newblock


\bibitem[Contributors(2023a)]%
        {esi2023document}
\bibfield{author}{\bibinfo{person}{CIRCT Contributors}.} \bibinfo{year}{2023}\natexlab{a}.
\newblock \bibinfo{booktitle}{\emph{CIRCT ESI Dialect}}.
\newblock
\urldef\tempurl%
\url{https://circt.llvm.org/docs/Dialects/ESI/}
\showURL{%
\tempurl}


\bibitem[Contributors(2023b)]%
        {handshake2023document}
\bibfield{author}{\bibinfo{person}{CIRCT Contributors}.} \bibinfo{year}{2023}\natexlab{b}.
\newblock \bibinfo{booktitle}{\emph{CIRCT Handshake Dialect}}.
\newblock
\urldef\tempurl%
\url{https://circt.llvm.org/docs/Dialects/Handshake/}
\showURL{%
\tempurl}


\bibitem[Contributors(2023c)]%
        {circt2023github}
\bibfield{author}{\bibinfo{person}{CIRCT Contributors}.} \bibinfo{year}{2023}\natexlab{c}.
\newblock \bibinfo{booktitle}{\emph{The CIRCT Project}}.
\newblock
\urldef\tempurl%
\url{https://github.com/llvm/circt}
\showURL{%
\tempurl}


\bibitem[Contributors(2023d)]%
        {iree2023github}
\bibfield{author}{\bibinfo{person}{IREE Contributors}.} \bibinfo{year}{2023}\natexlab{d}.
\newblock \bibinfo{booktitle}{\emph{The IREE Project}}.
\newblock
\urldef\tempurl%
\url{https://github.com/openxla/iree}
\showURL{%
\tempurl}


\bibitem[Contributors(2023e)]%
        {affine2023document}
\bibfield{author}{\bibinfo{person}{MLIR Contributors}.} \bibinfo{year}{2023}\natexlab{e}.
\newblock \bibinfo{booktitle}{\emph{MLIR Affine Dialect}}.
\newblock
\urldef\tempurl%
\url{https://mlir.llvm.org/docs/Dialects/Affine/}
\showURL{%
\tempurl}


\bibitem[Contributors(2023f)]%
        {linalg2023document}
\bibfield{author}{\bibinfo{person}{MLIR Contributors}.} \bibinfo{year}{2023}\natexlab{f}.
\newblock \bibinfo{booktitle}{\emph{MLIR LinAlg Dialect}}.
\newblock
\urldef\tempurl%
\url{https://mlir.llvm.org/docs/Dialects/Linalg/}
\showURL{%
\tempurl}


\bibitem[Contributors(2023g)]%
        {mlir2023github}
\bibfield{author}{\bibinfo{person}{MLIR Contributors}.} \bibinfo{year}{2023}\natexlab{g}.
\newblock \bibinfo{booktitle}{\emph{MLIR Project}}.
\newblock
\urldef\tempurl%
\url{https://github.com/llvm/llvm-project/tree/main/mlir}
\showURL{%
\tempurl}


\bibitem[Contributors(2023h)]%
        {tensor2023document}
\bibfield{author}{\bibinfo{person}{MLIR Contributors}.} \bibinfo{year}{2023}\natexlab{h}.
\newblock \bibinfo{booktitle}{\emph{MLIR Tensor Dialect}}.
\newblock
\urldef\tempurl%
\url{https://mlir.llvm.org/docs/Dialects/TensorOps/}
\showURL{%
\tempurl}


\bibitem[Contributors(2023i)]%
        {torchmlir2023github}
\bibfield{author}{\bibinfo{person}{Torch-MLIR Contributors}.} \bibinfo{year}{2023}\natexlab{i}.
\newblock \bibinfo{booktitle}{\emph{Torch-MLIR Project}}.
\newblock
\urldef\tempurl%
\url{https://github.com/llvm/torch-mlir/}
\showURL{%
\tempurl}


\bibitem[Cytron et~al\mbox{.}(1991)]%
        {cytron1991efficiently}
\bibfield{author}{\bibinfo{person}{Ron Cytron}, \bibinfo{person}{Jeanne Ferrante}, \bibinfo{person}{Barry~K Rosen}, \bibinfo{person}{Mark~N Wegman}, {and} \bibinfo{person}{F~Kenneth Zadeck}.} \bibinfo{year}{1991}\natexlab{}.
\newblock \showarticletitle{Efficiently computing static single assignment form and the control dependence graph}.
\newblock \bibinfo{journal}{\emph{ACM Transactions on Programming Languages and Systems (TOPLAS)}} \bibinfo{volume}{13}, \bibinfo{number}{4} (\bibinfo{year}{1991}), \bibinfo{pages}{451--490}.
\newblock


\bibitem[Du et~al\mbox{.}(2015)]%
        {du2015shidiannao}
\bibfield{author}{\bibinfo{person}{Zidong Du}, \bibinfo{person}{Robert Fasthuber}, \bibinfo{person}{Tianshi Chen}, \bibinfo{person}{Paolo Ienne}, \bibinfo{person}{Ling Li}, \bibinfo{person}{Tao Luo}, \bibinfo{person}{Xiaobing Feng}, \bibinfo{person}{Yunji Chen}, {and} \bibinfo{person}{Olivier Temam}.} \bibinfo{year}{2015}\natexlab{}.
\newblock \showarticletitle{ShiDianNao: Shifting vision processing closer to the sensor}. In \bibinfo{booktitle}{\emph{Proceedings of the 42nd Annual International Symposium on Computer Architecture}}. \bibinfo{pages}{92--104}.
\newblock


\bibitem[Durst et~al\mbox{.}(2020)]%
        {durst2020type}
\bibfield{author}{\bibinfo{person}{David Durst}, \bibinfo{person}{Matthew Feldman}, \bibinfo{person}{Dillon Huff}, \bibinfo{person}{David Akeley}, \bibinfo{person}{Ross Daly}, \bibinfo{person}{Gilbert~Louis Bernstein}, \bibinfo{person}{Marco Patrignani}, \bibinfo{person}{Kayvon Fatahalian}, {and} \bibinfo{person}{Pat Hanrahan}.} \bibinfo{year}{2020}\natexlab{}.
\newblock \showarticletitle{Type-directed scheduling of streaming accelerators}. In \bibinfo{booktitle}{\emph{Proceedings of the 41st ACM SIGPLAN Conference on Programming Language Design and Implementation}}. \bibinfo{pages}{408--422}.
\newblock


\bibitem[Ejjeh et~al\mbox{.}(2022)]%
        {ejjeh2022hpvm2fpga}
\bibfield{author}{\bibinfo{person}{Adel Ejjeh}, \bibinfo{person}{Leon Medvinsky}, \bibinfo{person}{Aaron Councilman}, \bibinfo{person}{Hemang Nehra}, \bibinfo{person}{Suraj Sharma}, \bibinfo{person}{Vikram Adve}, \bibinfo{person}{Luigi Nardi}, \bibinfo{person}{Eriko Nurvitadhi}, {and} \bibinfo{person}{Rob~A Rutenbar}.} \bibinfo{year}{2022}\natexlab{}.
\newblock \showarticletitle{HPVM2FPGA: Enabling true hardware-agnostic FPGA programming}. In \bibinfo{booktitle}{\emph{2022 IEEE 33rd International Conference on Application-specific Systems, Architectures and Processors (ASAP)}}. IEEE, \bibinfo{pages}{1--10}.
\newblock


\bibitem[Fahim et~al\mbox{.}(2021)]%
        {fahim2021hls4ml}
\bibfield{author}{\bibinfo{person}{Farah Fahim}, \bibinfo{person}{Benjamin Hawks}, \bibinfo{person}{Christian Herwig}, \bibinfo{person}{James Hirschauer}, \bibinfo{person}{Sergo Jindariani}, \bibinfo{person}{Nhan Tran}, \bibinfo{person}{Luca~P Carloni}, \bibinfo{person}{Giuseppe Di~Guglielmo}, \bibinfo{person}{Philip Harris}, \bibinfo{person}{Jeffrey Krupa}, {et~al\mbox{.}}} \bibinfo{year}{2021}\natexlab{}.
\newblock \showarticletitle{hls4ml: An open-source codesign workflow to empower scientific low-power machine learning devices}.
\newblock \bibinfo{journal}{\emph{arXiv preprint arXiv:2103.05579}} (\bibinfo{year}{2021}).
\newblock


\bibitem[Genc et~al\mbox{.}(2021)]%
        {genc2021gemmini}
\bibfield{author}{\bibinfo{person}{Hasan Genc}, \bibinfo{person}{Seah Kim}, \bibinfo{person}{Alon Amid}, \bibinfo{person}{Ameer Haj-Ali}, \bibinfo{person}{Vighnesh Iyer}, \bibinfo{person}{Pranav Prakash}, \bibinfo{person}{Jerry Zhao}, \bibinfo{person}{Daniel Grubb}, \bibinfo{person}{Harrison Liew}, \bibinfo{person}{Howard Mao}, {et~al\mbox{.}}} \bibinfo{year}{2021}\natexlab{}.
\newblock \showarticletitle{Gemmini: Enabling systematic deep-learning architecture evaluation via full-stack integration}. In \bibinfo{booktitle}{\emph{2021 58th ACM/IEEE Design Automation Conference (DAC)}}. IEEE, \bibinfo{pages}{769--774}.
\newblock


\bibitem[Guo et~al\mbox{.}(2022)]%
        {guo2022tapa}
\bibfield{author}{\bibinfo{person}{Licheng Guo}, \bibinfo{person}{Yuze Chi}, \bibinfo{person}{Jason Lau}, \bibinfo{person}{Linghao Song}, \bibinfo{person}{Xingyu Tian}, \bibinfo{person}{Moazin Khatti}, \bibinfo{person}{Weikang Qiao}, \bibinfo{person}{Jie Wang}, \bibinfo{person}{Ecenur Ustun}, \bibinfo{person}{Zhenman Fang}, {et~al\mbox{.}}} \bibinfo{year}{2022}\natexlab{}.
\newblock \showarticletitle{TAPA: A Scalable Task-Parallel Dataflow Programming Framework for Modern FPGAs with Co-Optimization of HLS and Physical Design}.
\newblock \bibinfo{journal}{\emph{arXiv preprint arXiv:2209.02663}} (\bibinfo{year}{2022}).
\newblock


\bibitem[He et~al\mbox{.}(2016)]%
        {he2016deep}
\bibfield{author}{\bibinfo{person}{Kaiming He}, \bibinfo{person}{Xiangyu Zhang}, \bibinfo{person}{Shaoqing Ren}, {and} \bibinfo{person}{Jian Sun}.} \bibinfo{year}{2016}\natexlab{}.
\newblock \showarticletitle{Deep residual learning for image recognition}. In \bibinfo{booktitle}{\emph{Proceedings of the IEEE conference on computer vision and pattern recognition}}. \bibinfo{pages}{770--778}.
\newblock


\bibitem[Howard et~al\mbox{.}(2017)]%
        {howard2017mobilenets}
\bibfield{author}{\bibinfo{person}{Andrew~G Howard}, \bibinfo{person}{Menglong Zhu}, \bibinfo{person}{Bo Chen}, \bibinfo{person}{Dmitry Kalenichenko}, \bibinfo{person}{Weijun Wang}, \bibinfo{person}{Tobias Weyand}, \bibinfo{person}{Marco Andreetto}, {and} \bibinfo{person}{Hartwig Adam}.} \bibinfo{year}{2017}\natexlab{}.
\newblock \showarticletitle{Mobilenets: Efficient convolutional neural networks for mobile vision applications}.
\newblock \bibinfo{journal}{\emph{arXiv preprint arXiv:1704.04861}} (\bibinfo{year}{2017}).
\newblock


\bibitem[Huang et~al\mbox{.}(2021)]%
        {huang2021pylog}
\bibfield{author}{\bibinfo{person}{Sitao Huang}, \bibinfo{person}{Kun Wu}, \bibinfo{person}{Hyunmin Jeong}, \bibinfo{person}{Chengyue Wang}, \bibinfo{person}{Deming Chen}, {and} \bibinfo{person}{Wen-mei Hwu}.} \bibinfo{year}{2021}\natexlab{}.
\newblock \showarticletitle{Pylog: An algorithm-centric python-based FPGA programming and synthesis flow}.
\newblock \bibinfo{journal}{\emph{IEEE Trans. Comput.}} \bibinfo{volume}{70}, \bibinfo{number}{12} (\bibinfo{year}{2021}), \bibinfo{pages}{2015--2028}.
\newblock


\bibitem[Ikarashi et~al\mbox{.}(2022)]%
        {ikarashi2022exocompilation}
\bibfield{author}{\bibinfo{person}{Yuka Ikarashi}, \bibinfo{person}{Gilbert~Louis Bernstein}, \bibinfo{person}{Alex Reinking}, \bibinfo{person}{Hasan Genc}, {and} \bibinfo{person}{Jonathan Ragan-Kelley}.} \bibinfo{year}{2022}\natexlab{}.
\newblock \showarticletitle{Exocompilation for productive programming of hardware accelerators}. In \bibinfo{booktitle}{\emph{Proceedings of the 43rd ACM SIGPLAN International Conference on Programming Language Design and Implementation}}. \bibinfo{pages}{703--718}.
\newblock


\bibitem[Inc(2022a)]%
        {vitishls2022userguide}
\bibfield{author}{\bibinfo{person}{Advanced Micro~Devices Inc}.} \bibinfo{year}{2022}\natexlab{a}.
\newblock \bibinfo{booktitle}{\emph{Vitis High-Level Synthesis User Guide UG1399 (v2022.2)}}.
\newblock


\bibitem[Inc(2022b)]%
        {intelhls2022userguide}
\bibfield{author}{\bibinfo{person}{Intel Inc}.} \bibinfo{year}{2022}\natexlab{b}.
\newblock \bibinfo{booktitle}{\emph{Intel High Level Synthesis Compiler Pro Edition Reference Manual (22.4)}}.
\newblock


\bibitem[Inc(2021)]%
        {legup2021document}
\bibfield{author}{\bibinfo{person}{Microchip~Technology Inc}.} \bibinfo{year}{2021}\natexlab{}.
\newblock \bibinfo{booktitle}{\emph{LegUp 2021.1 Documentation}}.
\newblock


\bibitem[Jo et~al\mbox{.}(2020)]%
        {jo2020soff}
\bibfield{author}{\bibinfo{person}{Gangwon Jo}, \bibinfo{person}{Heehoon Kim}, \bibinfo{person}{Jeesoo Lee}, {and} \bibinfo{person}{Jaejin Lee}.} \bibinfo{year}{2020}\natexlab{}.
\newblock \showarticletitle{SOFF: An OpenCL high-level synthesis framework for FPGAs}. In \bibinfo{booktitle}{\emph{2020 ACM/IEEE 47th Annual International Symposium on Computer Architecture (ISCA)}}. IEEE, \bibinfo{pages}{295--308}.
\newblock


\bibitem[Josipovic et~al\mbox{.}(2017)]%
        {josipovic2017out}
\bibfield{author}{\bibinfo{person}{Lana Josipovic}, \bibinfo{person}{Philip Brisk}, {and} \bibinfo{person}{Paolo Ienne}.} \bibinfo{year}{2017}\natexlab{}.
\newblock \showarticletitle{An out-of-order load-store queue for spatial computing}.
\newblock \bibinfo{journal}{\emph{ACM Transactions on Embedded Computing Systems (TECS)}} \bibinfo{volume}{16}, \bibinfo{number}{5s} (\bibinfo{year}{2017}), \bibinfo{pages}{1--19}.
\newblock


\bibitem[Josipovi{\'c} et~al\mbox{.}(2018)]%
        {josipovic2018dynamically}
\bibfield{author}{\bibinfo{person}{Lana Josipovi{\'c}}, \bibinfo{person}{Radhika Ghosal}, {and} \bibinfo{person}{Paolo Ienne}.} \bibinfo{year}{2018}\natexlab{}.
\newblock \showarticletitle{Dynamically scheduled high-level synthesis}. In \bibinfo{booktitle}{\emph{Proceedings of the 2018 ACM/SIGDA International Symposium on Field-Programmable Gate Arrays}}. \bibinfo{pages}{127--136}.
\newblock


\bibitem[Jouppi et~al\mbox{.}(2017)]%
        {jouppi2017datacenter}
\bibfield{author}{\bibinfo{person}{Norman~P Jouppi}, \bibinfo{person}{Cliff Young}, \bibinfo{person}{Nishant Patil}, \bibinfo{person}{David Patterson}, \bibinfo{person}{Gaurav Agrawal}, \bibinfo{person}{Raminder Bajwa}, \bibinfo{person}{Sarah Bates}, \bibinfo{person}{Suresh Bhatia}, \bibinfo{person}{Nan Boden}, \bibinfo{person}{Al Borchers}, {et~al\mbox{.}}} \bibinfo{year}{2017}\natexlab{}.
\newblock \showarticletitle{In-datacenter performance analysis of a tensor processing unit}. In \bibinfo{booktitle}{\emph{Proceedings of the 44th annual international symposium on computer architecture}}. \bibinfo{pages}{1--12}.
\newblock


\bibitem[Jun et~al\mbox{.}(2023)]%
        {jun2023autoscaledse}
\bibfield{author}{\bibinfo{person}{HyeGang Jun}, \bibinfo{person}{Hanchen Ye}, \bibinfo{person}{Hyunmin Jeong}, {and} \bibinfo{person}{Deming Chen}.} \bibinfo{year}{2023}\natexlab{}.
\newblock \showarticletitle{AutoScaleDSE: A Scalable Design Space Exploration Engine for High-Level Synthesis}.
\newblock \bibinfo{journal}{\emph{ACM Trans. Reconfigurable Technol. Syst.}} (\bibinfo{year}{2023}).
\newblock


\bibitem[Koeplinger et~al\mbox{.}(2018)]%
        {koeplinger2018spatial}
\bibfield{author}{\bibinfo{person}{David Koeplinger}, \bibinfo{person}{Matthew Feldman}, \bibinfo{person}{Raghu Prabhakar}, \bibinfo{person}{Yaqi Zhang}, \bibinfo{person}{Stefan Hadjis}, \bibinfo{person}{Ruben Fiszel}, \bibinfo{person}{Tian Zhao}, \bibinfo{person}{Luigi Nardi}, \bibinfo{person}{Ardavan Pedram}, \bibinfo{person}{Christos Kozyrakis}, {et~al\mbox{.}}} \bibinfo{year}{2018}\natexlab{}.
\newblock \showarticletitle{Spatial: A language and compiler for application accelerators}. In \bibinfo{booktitle}{\emph{Proceedings of the 39th ACM SIGPLAN Conference on Programming Language Design and Implementation}}. \bibinfo{pages}{296--311}.
\newblock


\bibitem[Lai et~al\mbox{.}(2019)]%
        {lai2019heterocl}
\bibfield{author}{\bibinfo{person}{Yi-Hsiang Lai}, \bibinfo{person}{Yuze Chi}, \bibinfo{person}{Yuwei Hu}, \bibinfo{person}{Jie Wang}, \bibinfo{person}{Cody~Hao Yu}, \bibinfo{person}{Yuan Zhou}, \bibinfo{person}{Jason Cong}, {and} \bibinfo{person}{Zhiru Zhang}.} \bibinfo{year}{2019}\natexlab{}.
\newblock \showarticletitle{HeteroCL: A multi-paradigm programming infrastructure for software-defined reconfigurable computing}. In \bibinfo{booktitle}{\emph{Proceedings of the 2019 ACM/SIGDA International Symposium on Field-Programmable Gate Arrays}}. \bibinfo{pages}{242--251}.
\newblock


\bibitem[Lattner and Adve(2004)]%
        {lattner2004llvm}
\bibfield{author}{\bibinfo{person}{Chris Lattner} {and} \bibinfo{person}{Vikram Adve}.} \bibinfo{year}{2004}\natexlab{}.
\newblock \showarticletitle{LLVM: A compilation framework for lifelong program analysis \& transformation}. In \bibinfo{booktitle}{\emph{International symposium on code generation and optimization, 2004. CGO 2004.}} IEEE, \bibinfo{pages}{75--86}.
\newblock


\bibitem[Lattner et~al\mbox{.}(2021)]%
        {lattner2021mlir}
\bibfield{author}{\bibinfo{person}{Chris Lattner}, \bibinfo{person}{Mehdi Amini}, \bibinfo{person}{Uday Bondhugula}, \bibinfo{person}{Albert Cohen}, \bibinfo{person}{Andy Davis}, \bibinfo{person}{Jacques Pienaar}, \bibinfo{person}{River Riddle}, \bibinfo{person}{Tatiana Shpeisman}, \bibinfo{person}{Nicolas Vasilache}, {and} \bibinfo{person}{Oleksandr Zinenko}.} \bibinfo{year}{2021}\natexlab{}.
\newblock \showarticletitle{MLIR: Scaling compiler infrastructure for domain specific computation}. In \bibinfo{booktitle}{\emph{2021 IEEE/ACM International Symposium on Code Generation and Optimization (CGO)}}. IEEE, \bibinfo{pages}{2--14}.
\newblock


\bibitem[Lattner et~al\mbox{.}(2020)]%
        {lattner2020mlir}
\bibfield{author}{\bibinfo{person}{Chris Lattner}, \bibinfo{person}{Jacques Pienaar}, \bibinfo{person}{Mehdi Amini}, \bibinfo{person}{Uday Bondhugula}, \bibinfo{person}{River Riddle}, \bibinfo{person}{Albert Cohen}, \bibinfo{person}{Tatiana Shpeisman}, \bibinfo{person}{Andy Davis}, \bibinfo{person}{Nicolas Vasilache}, {and} \bibinfo{person}{Oleksandr Zinenko}.} \bibinfo{year}{2020}\natexlab{}.
\newblock \showarticletitle{MLIR: A Compiler Infrastructure for the End of Moore's Law}.
\newblock \bibinfo{journal}{\emph{arXiv preprint arXiv:2002.11054}} (\bibinfo{year}{2020}).
\newblock


\bibitem[LeCun et~al\mbox{.}(1998)]%
        {lecun1998gradient}
\bibfield{author}{\bibinfo{person}{Yann LeCun}, \bibinfo{person}{L{\'e}on Bottou}, \bibinfo{person}{Yoshua Bengio}, {and} \bibinfo{person}{Patrick Haffner}.} \bibinfo{year}{1998}\natexlab{}.
\newblock \showarticletitle{Gradient-based learning applied to document recognition}.
\newblock \bibinfo{journal}{\emph{Proc. IEEE}} \bibinfo{volume}{86}, \bibinfo{number}{11} (\bibinfo{year}{1998}), \bibinfo{pages}{2278--2324}.
\newblock


\bibitem[Liu et~al\mbox{.}(2023)]%
        {liu2023unified}
\bibfield{author}{\bibinfo{person}{Qiaoyi Liu}, \bibinfo{person}{Jeff Setter}, \bibinfo{person}{Dillon Huff}, \bibinfo{person}{Maxwell Strange}, \bibinfo{person}{Kathleen Feng}, \bibinfo{person}{Mark Horowitz}, \bibinfo{person}{Priyanka Raina}, {and} \bibinfo{person}{Fredrik Kjolstad}.} \bibinfo{year}{2023}\natexlab{}.
\newblock \showarticletitle{Unified Buffer: Compiling Image Processing and Machine Learning Applications to Push-Memory Accelerators}.
\newblock \bibinfo{journal}{\emph{ACM Transactions on Architecture and Code Optimization}} \bibinfo{volume}{20}, \bibinfo{number}{2} (\bibinfo{year}{2023}), \bibinfo{pages}{1--26}.
\newblock


\bibitem[Margerm et~al\mbox{.}(2018)]%
        {margerm2018tapas}
\bibfield{author}{\bibinfo{person}{Steven Margerm}, \bibinfo{person}{Amirali Sharifian}, \bibinfo{person}{Apala Guha}, \bibinfo{person}{Arrvindh Shriraman}, {and} \bibinfo{person}{Gilles Pokam}.} \bibinfo{year}{2018}\natexlab{}.
\newblock \showarticletitle{TAPAS: Generating parallel accelerators from parallel programs}. In \bibinfo{booktitle}{\emph{2018 51st Annual IEEE/ACM International Symposium on Microarchitecture (MICRO)}}. IEEE, \bibinfo{pages}{245--257}.
\newblock


\bibitem[Moreau et~al\mbox{.}(2018)]%
        {moreau2018vta}
\bibfield{author}{\bibinfo{person}{Thierry Moreau}, \bibinfo{person}{Tianqi Chen}, \bibinfo{person}{Ziheng Jiang}, \bibinfo{person}{Luis Ceze}, \bibinfo{person}{Carlos Guestrin}, {and} \bibinfo{person}{Arvind Krishnamurthy}.} \bibinfo{year}{2018}\natexlab{}.
\newblock \showarticletitle{VTA: an open hardware-software stack for deep learning}.
\newblock \bibinfo{journal}{\emph{arXiv preprint arXiv:1807.04188}} (\bibinfo{year}{2018}).
\newblock


\bibitem[Moses et~al\mbox{.}(2021)]%
        {moses2021polygeist}
\bibfield{author}{\bibinfo{person}{William~S Moses}, \bibinfo{person}{Lorenzo Chelini}, \bibinfo{person}{Ruizhe Zhao}, {and} \bibinfo{person}{Oleksandr Zinenko}.} \bibinfo{year}{2021}\natexlab{}.
\newblock \showarticletitle{Polygeist: Raising C to polyhedral MLIR}. In \bibinfo{booktitle}{\emph{2021 30th International Conference on Parallel Architectures and Compilation Techniques (PACT)}}. IEEE, \bibinfo{pages}{45--59}.
\newblock


\bibitem[Mullapudi et~al\mbox{.}(2015)]%
        {mullapudi2015polymage}
\bibfield{author}{\bibinfo{person}{Ravi~Teja Mullapudi}, \bibinfo{person}{Vinay Vasista}, {and} \bibinfo{person}{Uday Bondhugula}.} \bibinfo{year}{2015}\natexlab{}.
\newblock \showarticletitle{Polymage: Automatic optimization for image processing pipelines}.
\newblock \bibinfo{journal}{\emph{ACM SIGARCH Computer Architecture News}} \bibinfo{volume}{43}, \bibinfo{number}{1} (\bibinfo{year}{2015}), \bibinfo{pages}{429--443}.
\newblock


\bibitem[Nardi et~al\mbox{.}(2019)]%
        {nardi2019practical}
\bibfield{author}{\bibinfo{person}{Luigi Nardi}, \bibinfo{person}{David Koeplinger}, {and} \bibinfo{person}{Kunle Olukotun}.} \bibinfo{year}{2019}\natexlab{}.
\newblock \showarticletitle{Practical design space exploration}. In \bibinfo{booktitle}{\emph{2019 IEEE 27th International Symposium on Modeling, Analysis, and Simulation of Computer and Telecommunication Systems (MASCOTS)}}. IEEE, \bibinfo{pages}{347--358}.
\newblock


\bibitem[Nigam et~al\mbox{.}(2020)]%
        {nigam2020predictable}
\bibfield{author}{\bibinfo{person}{Rachit Nigam}, \bibinfo{person}{Sachille Atapattu}, \bibinfo{person}{Samuel Thomas}, \bibinfo{person}{Zhijing Li}, \bibinfo{person}{Theodore Bauer}, \bibinfo{person}{Yuwei Ye}, \bibinfo{person}{Apurva Koti}, \bibinfo{person}{Adrian Sampson}, {and} \bibinfo{person}{Zhiru Zhang}.} \bibinfo{year}{2020}\natexlab{}.
\newblock \showarticletitle{Predictable accelerator design with time-sensitive affine types}. In \bibinfo{booktitle}{\emph{Proceedings of the 41st ACM SIGPLAN Conference on Programming Language Design and Implementation}}. \bibinfo{pages}{393--407}.
\newblock


\bibitem[Nigam et~al\mbox{.}(2021)]%
        {nigam2021compiler}
\bibfield{author}{\bibinfo{person}{Rachit Nigam}, \bibinfo{person}{Samuel Thomas}, \bibinfo{person}{Zhijing Li}, {and} \bibinfo{person}{Adrian Sampson}.} \bibinfo{year}{2021}\natexlab{}.
\newblock \showarticletitle{A compiler infrastructure for accelerator generators}. In \bibinfo{booktitle}{\emph{Proceedings of the 26th ACM International Conference on Architectural Support for Programming Languages and Operating Systems}}. \bibinfo{pages}{804--817}.
\newblock


\bibitem[Parashar et~al\mbox{.}(2019)]%
        {parashar2019timeloop}
\bibfield{author}{\bibinfo{person}{Angshuman Parashar}, \bibinfo{person}{Priyanka Raina}, \bibinfo{person}{Yakun~Sophia Shao}, \bibinfo{person}{Yu-Hsin Chen}, \bibinfo{person}{Victor~A Ying}, \bibinfo{person}{Anurag Mukkara}, \bibinfo{person}{Rangharajan Venkatesan}, \bibinfo{person}{Brucek Khailany}, \bibinfo{person}{Stephen~W Keckler}, {and} \bibinfo{person}{Joel Emer}.} \bibinfo{year}{2019}\natexlab{}.
\newblock \showarticletitle{Timeloop: A systematic approach to dnn accelerator evaluation}. In \bibinfo{booktitle}{\emph{2019 IEEE international symposium on performance analysis of systems and software (ISPASS)}}. IEEE, \bibinfo{pages}{304--315}.
\newblock


\bibitem[Paszke et~al\mbox{.}(2019)]%
        {paszke2019pytorch}
\bibfield{author}{\bibinfo{person}{Adam Paszke}, \bibinfo{person}{Sam Gross}, \bibinfo{person}{Francisco Massa}, \bibinfo{person}{Adam Lerer}, \bibinfo{person}{James Bradbury}, \bibinfo{person}{Gregory Chanan}, \bibinfo{person}{Trevor Killeen}, \bibinfo{person}{Zeming Lin}, \bibinfo{person}{Natalia Gimelshein}, \bibinfo{person}{Luca Antiga}, {et~al\mbox{.}}} \bibinfo{year}{2019}\natexlab{}.
\newblock \showarticletitle{Pytorch: An imperative style, high-performance deep learning library}.
\newblock \bibinfo{journal}{\emph{Advances in neural information processing systems}}  \bibinfo{volume}{32} (\bibinfo{year}{2019}).
\newblock


\bibitem[Pouchet et~al\mbox{.}(2012)]%
        {pouchet2012polybench}
\bibfield{author}{\bibinfo{person}{Louis-No{\"e}l Pouchet} {et~al\mbox{.}}} \bibinfo{year}{2012}\natexlab{}.
\newblock \showarticletitle{Polybench: The polyhedral benchmark suite}.
\newblock \bibinfo{journal}{\emph{URL: http://www. cs. ucla. edu/pouchet/software/polybench}}  \bibinfo{volume}{437} (\bibinfo{year}{2012}), \bibinfo{pages}{1--1}.
\newblock


\bibitem[Prabhakar et~al\mbox{.}(2017)]%
        {prabhakar2017plasticine}
\bibfield{author}{\bibinfo{person}{Raghu Prabhakar}, \bibinfo{person}{Yaqi Zhang}, \bibinfo{person}{David Koeplinger}, \bibinfo{person}{Matt Feldman}, \bibinfo{person}{Tian Zhao}, \bibinfo{person}{Stefan Hadjis}, \bibinfo{person}{Ardavan Pedram}, \bibinfo{person}{Christos Kozyrakis}, {and} \bibinfo{person}{Kunle Olukotun}.} \bibinfo{year}{2017}\natexlab{}.
\newblock \showarticletitle{Plasticine: A reconfigurable architecture for parallel paterns}.
\newblock \bibinfo{journal}{\emph{ACM SIGARCH Computer Architecture News}} \bibinfo{volume}{45}, \bibinfo{number}{2} (\bibinfo{year}{2017}), \bibinfo{pages}{389--402}.
\newblock


\bibitem[Redmon et~al\mbox{.}(2016)]%
        {redmon2016you}
\bibfield{author}{\bibinfo{person}{Joseph Redmon}, \bibinfo{person}{Santosh Divvala}, \bibinfo{person}{Ross Girshick}, {and} \bibinfo{person}{Ali Farhadi}.} \bibinfo{year}{2016}\natexlab{}.
\newblock \showarticletitle{You only look once: Unified, real-time object detection}. In \bibinfo{booktitle}{\emph{Proceedings of the IEEE conference on computer vision and pattern recognition}}. \bibinfo{pages}{779--788}.
\newblock


\bibitem[Rucker et~al\mbox{.}(2023)]%
        {rucker2023revet}
\bibfield{author}{\bibinfo{person}{Alexander Rucker}, \bibinfo{person}{Shiv Sundram}, \bibinfo{person}{Coleman Smith}, \bibinfo{person}{Matthew Vilim}, \bibinfo{person}{Raghu Prabhakar}, \bibinfo{person}{Fredrik Kjolstad}, {and} \bibinfo{person}{Kunle Olukotun}.} \bibinfo{year}{2023}\natexlab{}.
\newblock \showarticletitle{Revet: A Language and Compiler for Dataflow Threads}.
\newblock \bibinfo{journal}{\emph{arXiv preprint arXiv:2302.06124}} (\bibinfo{year}{2023}).
\newblock


\bibitem[Schardl et~al\mbox{.}(2017)]%
        {schardl2017tapir}
\bibfield{author}{\bibinfo{person}{Tao~B Schardl}, \bibinfo{person}{William~S Moses}, {and} \bibinfo{person}{Charles~E Leiserson}.} \bibinfo{year}{2017}\natexlab{}.
\newblock \showarticletitle{Tapir: Embedding fork-join parallelism into LLVM's intermediate representation}. In \bibinfo{booktitle}{\emph{Proceedings of the 22nd ACM SIGPLAN Symposium on Principles and Practice of Parallel Programming}}. \bibinfo{pages}{249--265}.
\newblock


\bibitem[Simonyan and Zisserman(2014)]%
        {simonyan2014very}
\bibfield{author}{\bibinfo{person}{Karen Simonyan} {and} \bibinfo{person}{Andrew Zisserman}.} \bibinfo{year}{2014}\natexlab{}.
\newblock \showarticletitle{Very deep convolutional networks for large-scale image recognition}.
\newblock \bibinfo{journal}{\emph{arXiv preprint arXiv:1409.1556}} (\bibinfo{year}{2014}).
\newblock


\bibitem[Sohrabizadeh et~al\mbox{.}(2022)]%
        {sohrabizadeh2022autodse}
\bibfield{author}{\bibinfo{person}{Atefeh Sohrabizadeh}, \bibinfo{person}{Cody~Hao Yu}, \bibinfo{person}{Min Gao}, {and} \bibinfo{person}{Jason Cong}.} \bibinfo{year}{2022}\natexlab{}.
\newblock \showarticletitle{AutoDSE: Enabling software programmers to design efficient FPGA accelerators}.
\newblock \bibinfo{journal}{\emph{ACM Transactions on Design Automation of Electronic Systems (TODAES)}} \bibinfo{volume}{27}, \bibinfo{number}{4} (\bibinfo{year}{2022}), \bibinfo{pages}{1--27}.
\newblock


\bibitem[Umuroglu et~al\mbox{.}(2017)]%
        {umuroglu2017finn}
\bibfield{author}{\bibinfo{person}{Yaman Umuroglu}, \bibinfo{person}{Nicholas~J Fraser}, \bibinfo{person}{Giulio Gambardella}, \bibinfo{person}{Michaela Blott}, \bibinfo{person}{Philip Leong}, \bibinfo{person}{Magnus Jahre}, {and} \bibinfo{person}{Kees Vissers}.} \bibinfo{year}{2017}\natexlab{}.
\newblock \showarticletitle{Finn: A framework for fast, scalable binarized neural network inference}. In \bibinfo{booktitle}{\emph{Proceedings of the 2017 ACM/SIGDA international symposium on field-programmable gate arrays}}. \bibinfo{pages}{65--74}.
\newblock


\bibitem[Wang et~al\mbox{.}(2013)]%
        {wang2013memory}
\bibfield{author}{\bibinfo{person}{Yuxin Wang}, \bibinfo{person}{Peng Li}, \bibinfo{person}{Peng Zhang}, \bibinfo{person}{Chen Zhang}, {and} \bibinfo{person}{Jason Cong}.} \bibinfo{year}{2013}\natexlab{}.
\newblock \showarticletitle{Memory partitioning for multidimensional arrays in high-level synthesis}. In \bibinfo{booktitle}{\emph{Proceedings of the 50th Annual Design Automation Conference}}. \bibinfo{pages}{1--8}.
\newblock


\bibitem[Wei et~al\mbox{.}(2018)]%
        {wei2018tgpa}
\bibfield{author}{\bibinfo{person}{Xuechao Wei}, \bibinfo{person}{Yun Liang}, \bibinfo{person}{Xiuhong Li}, \bibinfo{person}{Cody~Hao Yu}, \bibinfo{person}{Peng Zhang}, {and} \bibinfo{person}{Jason Cong}.} \bibinfo{year}{2018}\natexlab{}.
\newblock \showarticletitle{TGPA: Tile-grained pipeline architecture for low latency CNN inference}. In \bibinfo{booktitle}{\emph{2018 IEEE/ACM International Conference on Computer-Aided Design (ICCAD)}}. ACM, \bibinfo{pages}{1--8}.
\newblock


\bibitem[Xiang et~al\mbox{.}(2022)]%
        {xiang2022heteroflow}
\bibfield{author}{\bibinfo{person}{Shaojie Xiang}, \bibinfo{person}{Yi-Hsiang Lai}, \bibinfo{person}{Yuan Zhou}, \bibinfo{person}{Hongzheng Chen}, \bibinfo{person}{Niansong Zhang}, \bibinfo{person}{Debjit Pal}, {and} \bibinfo{person}{Zhiru Zhang}.} \bibinfo{year}{2022}\natexlab{}.
\newblock \showarticletitle{Heteroflow: An accelerator programming model with decoupled data placement for software-defined fpgas}. In \bibinfo{booktitle}{\emph{Proceedings of the 2022 ACM/SIGDA International Symposium on Field-Programmable Gate Arrays}}. \bibinfo{pages}{78--88}.
\newblock


\bibitem[Xu et~al\mbox{.}(2022)]%
        {xu2022hector}
\bibfield{author}{\bibinfo{person}{Ruifan Xu}, \bibinfo{person}{Youwei Xiao}, \bibinfo{person}{Jin Luo}, {and} \bibinfo{person}{Yun Liang}.} \bibinfo{year}{2022}\natexlab{}.
\newblock \showarticletitle{HECTOR: A Multi-Level Intermediate Representation for Hardware Synthesis Methodologies}. In \bibinfo{booktitle}{\emph{Proceedings of the 41st IEEE/ACM International Conference on Computer-Aided Design}}. \bibinfo{pages}{1--9}.
\newblock


\bibitem[Ye et~al\mbox{.}(2022a)]%
        {hpca2022scalehls}
\bibfield{author}{\bibinfo{person}{Hanchen Ye}, \bibinfo{person}{Cong Hao}, \bibinfo{person}{Jianyi Cheng}, \bibinfo{person}{Hyunmin Jeong}, \bibinfo{person}{Jack Huang}, \bibinfo{person}{Stephen Neuendorffer}, {and} \bibinfo{person}{Deming Chen}.} \bibinfo{year}{2022}\natexlab{a}.
\newblock \showarticletitle{Scalehls: A new scalable high-level synthesis framework on multi-level intermediate representation}. In \bibinfo{booktitle}{\emph{2022 IEEE International Symposium on High-Performance Computer Architecture (HPCA)}}. IEEE, \bibinfo{pages}{741--755}.
\newblock


\bibitem[Ye et~al\mbox{.}(2022b)]%
        {dac2022scalehls}
\bibfield{author}{\bibinfo{person}{Hanchen Ye}, \bibinfo{person}{HyeGang Jun}, \bibinfo{person}{Hyunmin Jeong}, \bibinfo{person}{Stephen Neuendorffer}, {and} \bibinfo{person}{Deming Chen}.} \bibinfo{year}{2022}\natexlab{b}.
\newblock \showarticletitle{ScaleHLS: a scalable high-level synthesis framework with multi-level transformations and optimizations}. In \bibinfo{booktitle}{\emph{Proceedings of the 59th ACM/IEEE Design Automation Conference}}. \bibinfo{pages}{1355--1358}.
\newblock


\bibitem[Ye et~al\mbox{.}(2020)]%
        {ye2020hybriddnn}
\bibfield{author}{\bibinfo{person}{Hanchen Ye}, \bibinfo{person}{Xiaofan Zhang}, \bibinfo{person}{Zhize Huang}, \bibinfo{person}{Gengsheng Chen}, {and} \bibinfo{person}{Deming Chen}.} \bibinfo{year}{2020}\natexlab{}.
\newblock \showarticletitle{HybridDNN: A framework for high-performance hybrid DNN accelerator design and implementation}. In \bibinfo{booktitle}{\emph{2020 57th ACM/IEEE Design Automation Conference (DAC)}}. IEEE, \bibinfo{pages}{1--6}.
\newblock


\bibitem[Yu et~al\mbox{.}(2021)]%
        {yu2021chimera}
\bibfield{author}{\bibinfo{person}{Mang Yu}, \bibinfo{person}{Sitao Huang}, {and} \bibinfo{person}{Deming Chen}.} \bibinfo{year}{2021}\natexlab{}.
\newblock \showarticletitle{Chimera: A hybrid machine learning-driven multi-objective design space exploration tool for fpga high-level synthesis}. In \bibinfo{booktitle}{\emph{Intelligent Data Engineering and Automated Learning--IDEAL 2021: 22nd International Conference, IDEAL 2021, Manchester, UK, November 25--27, 2021, Proceedings 22}}. Springer, \bibinfo{pages}{524--536}.
\newblock


\bibitem[Zeiler and Fergus(2014)]%
        {zeiler2014visualizing}
\bibfield{author}{\bibinfo{person}{Matthew~D Zeiler} {and} \bibinfo{person}{Rob Fergus}.} \bibinfo{year}{2014}\natexlab{}.
\newblock \showarticletitle{Visualizing and understanding convolutional networks}. In \bibinfo{booktitle}{\emph{Computer Vision--ECCV 2014: 13th European Conference, Zurich, Switzerland, September 6-12, 2014, Proceedings, Part I 13}}. Springer, \bibinfo{pages}{818--833}.
\newblock


\bibitem[Zhang et~al\mbox{.}(2021)]%
        {zhang2021boostgcn}
\bibfield{author}{\bibinfo{person}{Bingyi Zhang}, \bibinfo{person}{Rajgopal Kannan}, {and} \bibinfo{person}{Viktor Prasanna}.} \bibinfo{year}{2021}\natexlab{}.
\newblock \showarticletitle{Boostgcn: A framework for optimizing gcn inference on fpga}. In \bibinfo{booktitle}{\emph{2021 IEEE 29th Annual International Symposium on Field-Programmable Custom Computing Machines (FCCM)}}. IEEE, \bibinfo{pages}{29--39}.
\newblock


\bibitem[Zhang et~al\mbox{.}(2015)]%
        {zhang2015optimizing}
\bibfield{author}{\bibinfo{person}{Chen Zhang}, \bibinfo{person}{Peng Li}, \bibinfo{person}{Guangyu Sun}, \bibinfo{person}{Yijin Guan}, \bibinfo{person}{Bingjun Xiao}, {and} \bibinfo{person}{Jason Cong}.} \bibinfo{year}{2015}\natexlab{}.
\newblock \showarticletitle{Optimizing FPGA-based accelerator design for deep convolutional neural networks}. In \bibinfo{booktitle}{\emph{Proceedings of the 2015 ACM/SIGDA international symposium on field-programmable gate arrays}}. \bibinfo{pages}{161--170}.
\newblock


\bibitem[Zhang et~al\mbox{.}(2018)]%
        {zhang2018dnnbuilder}
\bibfield{author}{\bibinfo{person}{Xiaofan Zhang}, \bibinfo{person}{Junsong Wang}, \bibinfo{person}{Chao Zhu}, \bibinfo{person}{Yonghua Lin}, \bibinfo{person}{Jinjun Xiong}, \bibinfo{person}{Wen-mei Hwu}, {and} \bibinfo{person}{Deming Chen}.} \bibinfo{year}{2018}\natexlab{}.
\newblock \showarticletitle{DNNBuilder: An automated tool for building high-performance DNN hardware accelerators for FPGAs}. In \bibinfo{booktitle}{\emph{2018 IEEE/ACM International Conference on Computer-Aided Design (ICCAD)}}. ACM, \bibinfo{pages}{1--8}.
\newblock


\bibitem[Zhang et~al\mbox{.}(2020)]%
        {zhang2020dnnexplorer}
\bibfield{author}{\bibinfo{person}{Xiaofan Zhang}, \bibinfo{person}{Hanchen Ye}, \bibinfo{person}{Junsong Wang}, \bibinfo{person}{Yonghua Lin}, \bibinfo{person}{Jinjun Xiong}, \bibinfo{person}{Wen-mei Hwu}, {and} \bibinfo{person}{Deming Chen}.} \bibinfo{year}{2020}\natexlab{}.
\newblock \showarticletitle{DNNExplorer: a framework for modeling and exploring a novel paradigm of FPGA-based DNN accelerator}. In \bibinfo{booktitle}{\emph{Proceedings of the 39th International Conference on Computer-Aided Design}}. \bibinfo{pages}{1--9}.
\newblock


\bibitem[Zhao et~al\mbox{.}(2017)]%
        {zhao2017comba}
\bibfield{author}{\bibinfo{person}{Jieru Zhao}, \bibinfo{person}{Liang Feng}, \bibinfo{person}{Sharad Sinha}, \bibinfo{person}{Wei Zhang}, \bibinfo{person}{Yun Liang}, {and} \bibinfo{person}{Bingsheng He}.} \bibinfo{year}{2017}\natexlab{}.
\newblock \showarticletitle{COMBA: A comprehensive model-based analysis framework for high level synthesis of real applications}. In \bibinfo{booktitle}{\emph{2017 IEEE/ACM International Conference on Computer-Aided Design (ICCAD)}}. IEEE, \bibinfo{pages}{430--437}.
\newblock


\bibitem[Zhao et~al\mbox{.}(2022)]%
        {zhao2022polsca}
\bibfield{author}{\bibinfo{person}{Ruizhe Zhao}, \bibinfo{person}{Jianyi Cheng}, \bibinfo{person}{Wayne Luk}, {and} \bibinfo{person}{George~A Constantinides}.} \bibinfo{year}{2022}\natexlab{}.
\newblock \showarticletitle{POLSCA: Polyhedral High-Level Synthesis with Compiler Transformations}. In \bibinfo{booktitle}{\emph{32nd International Conference on Field Programmable Logic and Applications (FPL'22)}}.
\newblock


\bibitem[Zhao et~al\mbox{.}(2023)]%
        {zhao2023sigma}
\bibfield{author}{\bibinfo{person}{Tian Zhao}, \bibinfo{person}{Alexander Rucker}, {and} \bibinfo{person}{Kunle Olukotun}.} \bibinfo{year}{2023}\natexlab{}.
\newblock \showarticletitle{Sigma: Compiling Einstein Summations to Locality-Aware Dataflow}. In \bibinfo{booktitle}{\emph{Proceedings of the 28th ACM International Conference on Architectural Support for Programming Languages and Operating Systems, Volume 2}}. \bibinfo{pages}{718--732}.
\newblock


\bibitem[Zhao et~al\mbox{.}(2018)]%
        {zhao2018deepthings}
\bibfield{author}{\bibinfo{person}{Zhuoran Zhao}, \bibinfo{person}{Kamyar~Mirzazad Barijough}, {and} \bibinfo{person}{Andreas Gerstlauer}.} \bibinfo{year}{2018}\natexlab{}.
\newblock \showarticletitle{Deepthings: Distributed adaptive deep learning inference on resource-constrained iot edge clusters}.
\newblock \bibinfo{journal}{\emph{IEEE Transactions on Computer-Aided Design of Integrated Circuits and Systems}} \bibinfo{volume}{37}, \bibinfo{number}{11} (\bibinfo{year}{2018}), \bibinfo{pages}{2348--2359}.
\newblock


\bibitem[Zhong et~al\mbox{.}(2016)]%
        {zhong2016lin}
\bibfield{author}{\bibinfo{person}{Guanwen Zhong}, \bibinfo{person}{Alok Prakash}, \bibinfo{person}{Yun Liang}, \bibinfo{person}{Tulika Mitra}, {and} \bibinfo{person}{Smail Niar}.} \bibinfo{year}{2016}\natexlab{}.
\newblock \showarticletitle{Lin-analyzer: A high-level performance analysis tool for FPGA-based accelerators}. In \bibinfo{booktitle}{\emph{Proceedings of the 53rd Annual Design Automation Conference}}. \bibinfo{pages}{1--6}.
\newblock


\bibitem[Zhou et~al\mbox{.}(2021)]%
        {zhou2021mocha}
\bibfield{author}{\bibinfo{person}{Peipei Zhou}, \bibinfo{person}{Jiayi Sheng}, \bibinfo{person}{Cody~Hao Yu}, \bibinfo{person}{Peng Wei}, \bibinfo{person}{Jie Wang}, \bibinfo{person}{Di Wu}, {and} \bibinfo{person}{Jason Cong}.} \bibinfo{year}{2021}\natexlab{}.
\newblock \showarticletitle{Mocha: Multinode cost optimization in heterogeneous clouds with accelerators}. In \bibinfo{booktitle}{\emph{The 2021 ACM/SIGDA International Symposium on Field-Programmable Gate Arrays}}. \bibinfo{pages}{273--279}.
\newblock


\bibitem[Zhuang et~al\mbox{.}(2023)]%
        {zhuang2023charm}
\bibfield{author}{\bibinfo{person}{Jinming Zhuang}, \bibinfo{person}{Jason Lau}, \bibinfo{person}{Hanchen Ye}, \bibinfo{person}{Zhuoping Yang}, \bibinfo{person}{Yubo Du}, \bibinfo{person}{Jack Lo}, \bibinfo{person}{Kristof Denolf}, \bibinfo{person}{Stephen Neuendorffer}, \bibinfo{person}{Alex Jones}, \bibinfo{person}{Jingtong Hu}, {et~al\mbox{.}}} \bibinfo{year}{2023}\natexlab{}.
\newblock \showarticletitle{CHARM: Composing Heterogeneous AcceleRators for Matrix Multiply on Versal ACAP Architecture}.
\newblock \bibinfo{journal}{\emph{arXiv preprint arXiv:2301.02359}} (\bibinfo{year}{2023}).
\newblock


\bibitem[Zuo et~al\mbox{.}(2015)]%
        {zuo2015polyhedral}
\bibfield{author}{\bibinfo{person}{Wei Zuo}, \bibinfo{person}{Warren Kemmerer}, \bibinfo{person}{Jong~Bin Lim}, \bibinfo{person}{Louis-No{\"e}l Pouchet}, \bibinfo{person}{Andrey Ayupov}, \bibinfo{person}{Taemin Kim}, \bibinfo{person}{Kyungtae Han}, {and} \bibinfo{person}{Deming Chen}.} \bibinfo{year}{2015}\natexlab{}.
\newblock \showarticletitle{A polyhedral-based systemc modeling and generation framework for effective low-power design space exploration}. In \bibinfo{booktitle}{\emph{2015 IEEE/ACM International Conference on Computer-Aided Design (ICCAD)}}. IEEE, \bibinfo{pages}{357--364}.
\newblock


\bibitem[Zuo et~al\mbox{.}(2017)]%
        {zuo2017accurate}
\bibfield{author}{\bibinfo{person}{Wei Zuo}, \bibinfo{person}{Louis-Noel Pouchet}, \bibinfo{person}{Andrey Ayupov}, \bibinfo{person}{Taemin Kim}, \bibinfo{person}{Chung-Wei Lin}, \bibinfo{person}{Shinichi Shiraishi}, {and} \bibinfo{person}{Deming Chen}.} \bibinfo{year}{2017}\natexlab{}.
\newblock \showarticletitle{Accurate high-level modeling and automated hardware/software co-design for effective SoC design space exploration}. In \bibinfo{booktitle}{\emph{Proceedings of the 54th Annual Design Automation Conference 2017}}. \bibinfo{pages}{1--6}.
\newblock


\end{thebibliography}


\end{document}